\documentclass[aps,twocolumn,prd,showpacs,showkeys,preprintnumbers,superscriptaddress,nobibnotes,floatfix,longbibliography,notitlepage,nofootinbib]{revtex4-2}

\pdfoutput=1
\usepackage{amsmath}
\usepackage{amsfonts}
\usepackage{amssymb}
\usepackage{mathrsfs}
\usepackage{color}
\usepackage{slashed}
\usepackage{graphicx}
\usepackage{xcolor}
\usepackage[colorlinks=true,allcolors=blue]{hyperref}
\usepackage{multirow}
\usepackage{upgreek}
\usepackage[capitalise]{cleveref}
\usepackage[colorinlistoftodos]{todonotes}

\usepackage{siunitx}

\DeclareSIUnit \s {\second}
\DeclareSIUnit \ns {\nano\second}
\DeclareSIUnit \mus {\micro\second}
\DeclareSIUnit \ms {\milli\second}
\DeclareSIUnit \MB {\mega\byte}
\DeclareSIUnit \GB {\giga\byte}
\DeclareSIUnit \TB {\tera\byte}
\DeclareSIUnit \PB {\peta\byte}
\DeclareSIUnit \Mbps {\mega\bit/\s}
\DeclareSIUnit \Gbps {\giga\bit/\s}
\DeclareSIUnit \Tbps {\tera\bit/\s}
\DeclareSIUnit \Pbps {\peta\bit/\s}
\DeclareSIUnit \kton {\kilo\tonne} 
\DeclareSIUnit \kt {\kilo\tonne}
\DeclareSIUnit \kty {\kilo\tonne-\year}
\DeclareSIUnit \Mt {\mega\tonne}
\DeclareSIUnit \eV {\electronvolt}
\DeclareSIUnit \keV {\kilo\electronvolt}
\DeclareSIUnit \MeV {\mega\electronvolt}
\DeclareSIUnit \GeV {\giga\electronvolt}
\DeclareSIUnit \TeV {\tera\electronvolt}
\DeclareSIUnit \PeV {\peta\electronvolt}
\DeclareSIUnit \EeV {\exa\electronvolt}
\DeclareSIUnit \m {\meter}
\DeclareSIUnit \cm {\centi\meter}
\DeclareSIUnit \nm {\nano\meter}
\DeclareSIUnit \in {\inchcommand}
\DeclareSIUnit \km {\kilo\meter}
\DeclareSIUnit \kV {\kilo\volt}
\DeclareSIUnit \kW {\kilo\watt}
\DeclareSIUnit \MW {\mega\watt}
\DeclareSIUnit \MHz {\mega\hertz}
\DeclareSIUnit \mrad {\milli\radian}
\DeclareSIUnit \year {years}
\DeclareSIUnit \POT {POT}
\DeclareSIUnit \sig {$\sigma$}
\DeclareSIUnit\parsec{pc}
\DeclareSIUnit\lightyear{ly}
\DeclareSIUnit\foot{ft}
\DeclareSIUnit\ft{ft}
\DeclareSIUnit \ppb{ppb}
\DeclareSIUnit \ppt{ppt}
\DeclareSIUnit \samples{S}
\DeclareSIUnit \pe{PE}
\DeclareSIUnit \GeVmwe{GeV/mwe}
\DeclareSIUnit \mwe{mwe}

\newcommand{\enu}{\E_\enu}

\usepackage{tikz}
\usepackage{tikz-feynman}

\usetikzlibrary{trees}
\usetikzlibrary{decorations.pathmorphing}
\usetikzlibrary{decorations.markings}

\definecolor{kyelloworange}   {RGB}{255, 210,  110}
\definecolor{jblue}  {RGB}{20,50,100}
\definecolor{npurple}  {RGB} {153, 51, 204}
\definecolor{wred}   {RGB}{217,0,56}
\definecolor{white}   {RGB}{255,255,255}
  
\definecolor{korange}   {RGB}{235, 80,  43}
\definecolor{korange2}   {RGB}{245, 100,  63}
\definecolor{kyelloworange}   {RGB}{255, 210,  110}
\definecolor{kyelloworange2}   {RGB}{240, 170,  90}
\definecolor{kred}   {RGB}{204,  102, 153}
\definecolor{kpurple}   {RGB}{153,  61, 190}
\definecolor{kpurplelight}   {RGB}{213,  161, 230}

\tikzset{
	    gaugeboson/.style={decorate,decoration={snake, amplitude = 3pt, post length = 1 pt, pre length = 4 pt},draw=magenta},
	    fermion/.style={draw=black,postaction={decorate},decoration={markings,mark=at position .55}},
	    fermionin/.style={draw=black,postaction={decorate},decoration={markings,mark=at position .55 with {\arrow[draw=black]{<}}}},
	    fermionout/.style={draw=black,postaction={decorate},decoration={markings,mark=at position .55 with {\arrow[draw=black]{>}}}},
	    gluon/.style={decorate,draw=magenta,decoration={coil,amplitude = 6pt,segment length=8pt}},
	    connect/.style={draw=black,postaction={decorate},decoration={markings}}, 
	    gluon2/.style={decorate,draw=magenta,decoration={coil,amplitude = 3pt,segment length=4pt}},
	    gaugeboson2/.style={decorate,decoration={snake, amplitude = 3pt, segment length=6pt},draw=black}
	}
	
\tikzset{
	  photon/.style={decorate, decoration={snake}, draw=npurple,very thick},
	  boson/.style={decorate, decoration={snake}, draw=npurple,very thick},
	  electron/.style={draw=jblue,very thick, postaction={decorate},
	           decoration={markings,mark=at position .55 with {\arrow[draw=jblue]{>}}}
	  },
	  electron2/.style={draw=jblue,very thick, postaction={decorate},
	           decoration={markings,mark=at position .55 with {\arrow[draw=jblue]{<}}}
	  },
	  fermion/.style={draw=jblue,very thick, postaction={decorate},
	            decoration={markings,mark=at position .55 with {\arrow[draw=jblue]{}}}
	  },
	  gluon/.style={decorate, draw=korange,very thick, 
	    decoration={coil,amplitude=4pt, segment length=6pt}},
	  higgs/.style={draw=wred,very thick, postaction={decorate},
	           decoration={markings,mark=at position .55 with {\arrow[draw=wred]{>}}}
	  },
	  nothing/.style={draw=white,very thick}
}

\begin{document}

\title{Explaining the MiniBooNE Excess Through a Mixed Model of Oscillation and Decay}

\author{S.~Vergani}
\email{sv408@hep.phy.cam.ac.uk}
\affiliation{University of Cambridge, Cambridge CB3 0HE, United Kingdom}

\author{N.W.~Kamp}
\email{nwkamp@mit.edu}
\affiliation{Dept.~of Physics, Massachusetts Institute of Technology, Cambridge, MA 02139, USA}

\author{A.~Diaz}
\email{diaza@mit.edu}
\affiliation{Dept.~of Physics, Massachusetts Institute of Technology, Cambridge, MA 02139, USA}

\author{C.A.~Arg{\"u}elles}
\email{carguelles@fas.harvard.edu}
\affiliation{Dept.~of Physics, Harvard University, Cambridge, MA 02138, USA}

\author{J.M.~Conrad}
\email{conrad@mit.edu}
\affiliation{Dept.~of Physics, Massachusetts Institute of Technology, Cambridge, MA 02139, USA}

\author{M.H.~Shaevitz}
\email{shaevitz@nevis.columbia.edu}
\affiliation{Dept.~of Physics, Columbia University, New York, NY, 10027, USA}

\author{M.A.~Uchida}
\email{mauchida@hep.phy.cam.ac.uk}
\affiliation{University of Cambridge, Cambridge CB3 0HE, United Kingdom}

\date{\today}

\begin{abstract}

The electron-like excess observed by the MiniBooNE experiment is explained with a model comprising a new low mass state ($\mathcal{O}(1)$ eV) participating in neutrino oscillations and a new high mass state ($\mathcal{O}(100)$ MeV) that decays to $\nu+\gamma$.
Short-baseline oscillation data sets are used to predict the oscillation parameters.  
Fitting the MiniBooNE energy and scattering angle data, there is a narrow joint allowed region for the decay contribution at 95\% CL.
The result is a substantial improvement over the single sterile neutrino oscillation model, with $\Delta \chi^2/dof$ = 19.3/2 for a decay coupling of $2.8 \times 10^{-7}$ GeV$^{-1}$, high mass state of 376 MeV, oscillation mixing angle of $7\times 10^{-4}$ and mass splitting of $1.3$ eV$^2$.
This model predicts that no clear oscillation signature will be observed in the FNAL short baseline program due to the low signal-level.

\end{abstract}

\maketitle

\section*{Introduction}
For the past 25 years, anomalies have been observed in short-baseline (SBL) neutrino oscillation experiments.
These have been studied under a model called ``3+1'' that introduces a new non-interacting, hence ``sterile,'' state with mass of $\mathcal{O}(1\,\rm{eV})$, in addition to the three Standard Model (SM) neutrino states.
In this model, $\nu_\mu \rightarrow \nu_e$ appearance,  $\nu_e$ disappearance, and $\nu_\mu$ disappearance searches should all point to neutrino oscillations at $L/E \sim 1$~m/MeV, where $L$ is the distance a neutrino of energy $E$ travels, with a consistent set of flavor mixing parameters~\cite{Collin:2016aqd,Dentler:2018sju,Diaz:2019fwt,Boser:2019rta}.
However, while individually the data appear to fit oscillations, global fits find a small probability that all of the relevant data sets are explained by the same parameters~\cite{Dentler:2018sju,Diaz:2019fwt}, as measured by the Parameter Goodness of Fit (PGF) test~\cite{Maltoni:2001bc,Maltoni:2003cu}.
In particular, appearance data from MiniBooNE produces large tension between appearance and disappearance in the 3+1 model.
This is because the 3+1 best-fit parameters from the other data sets yield a poor fit to the lowest energy range of the MiniBooNE anomaly~\cite{Giunti:2013aea}. 
Therefore, there is significant interest in explanations for MiniBooNE beyond the 3+1 model; for example, one can consider decays of a sterile neutrinos into active neutrinos and singlet scalars~\cite{Dentler:2019dhz,deGouvea:2019qre}.

The MiniBooNE anomaly is a 4.8$\sigma$ excess of electron-like events observed in interactions from a predominantly muon neutrino beam in a Cherenkov detector~\cite{MiniBooNE:2020pnu}, which cannot distinguish between electromagnetic showers from electrons and photons.
Hence, a favored alternative to the 3+1 model has been to introduce MeV-scale heavy neutral leptons (HNLs) that decay via $\mathcal{N} \rightarrow \nu \gamma$ within the detector, where the photon is then misidentified as an electron~\cite{Gninenko:2009ks,McKeen:2010rx,Dib:2011jh,Gninenko:2012rw,Masip:2012ke,Ballett:2016opr,Magill:2018jla,Fischer:2019fbw}; see~Refs.\cite{Bertuzzo:2018ftf,Bertuzzo:2018itn,Ballett:2018ynz,Arguelles:2018mtc,Ballett:2019cqp,Ballett:2019pyw,Abdullahi:2020nyr,Abdallah:2020vgg,Abdallah:2020biq} for misidentified di-electron scenarios.
These initial studies of $\mathcal{N}$-decay models describe the MiniBooNE energy distribution well but omit the 3+1 oscillations predicted from fits to the other anomalies.

In this work, we explore a combination of the two explanations by fitting the MiniBooNE energy and angle distributions using a combined model, 3+1+$\mathcal{N}$-decay.
The 3+1 oscillation component has been obtained by fitting SBL data sets other than MiniBooNE appearance.
We will show that such a model explains the data well, identifying a highly limited range for the four model parameters: the mixing angle, $\sin^2 2\theta$, and mass splitting, $\Delta m^2$, for the oscillation; and the HNL mass, $m_{\mathcal{N}}$, and photon coupling, $d$, for the decay.

\section*{Model}
The combination of eV-scale and MeV-scale mass states is motivated if the two are members of a family of $\mathcal{N}_j$ where $j=1,2,3$.
If the mass splittings are similar to the quark and charged-lepton sectors, then the family might also include a keV-scale member~\cite{Adhikari:2016bei,non_minimal_hnl_loi}.
All members may interact with photons at a weak level through a dipole portal interaction~\cite{Magill:2018jla}, also known as neutrino magnetic moment~\cite{Fujikawa:1980yx,Pal:1981rm,Shrock:1982sc,Dvornikov:2003js,Giunti:2014ixa,Brdar:2020quo}.
Thus, the $\mathcal{N}_1=\nu_4$ can decay, but the lifetime is typically longer than the age of the Universe~\cite{Pal:1981rm,Nieves:1982bq}.
The keV-scale mass state, $\mathcal{N}_2$, would have a lifetime that is too long to be observed through decay in terrestrial experiments but could explain observed X-ray lines~\cite{Adhikari:2016bei,Abazajian:2017tcc}.
Only the $\mathcal{N} \equiv \mathcal{N}_3$ would decay on length scales relevant to SBL experiments.
Conversely, only the eV-scale mass state would have sufficiently small mass splitting with respect to the light neutrino states~\cite{Tanabashi:2018oca,Esteban:2018azc,deSalas:2017kay,Capozzi:2016rtj} to form observable oscillations.  
In the 3+1+$\mathcal{N}$-decay model, any given SBL experiment may be sensitive to signatures of 3+1 oscillations only, $\mathcal{N} \rightarrow \nu \gamma$ only, or both.

\begin{table}[t]
    \centering
    \begin{tabular}{|l|c|c|}
    \hline
    Used to Test & References (Flux Type) & Type of Fit \\
    \hline
    $\bar \nu_e$ disappearance & \cite{Bugey, NEOS, DANSS, PhysRevD.103.032001, PhysRevD.102.052002} (Reactor)& \\
    $\nu_e$ disappearance & \cite{ConradShaevitz,Gallex, SAGE} (Source) &  \\
    $\bar \nu_\mu \rightarrow \bar \nu_e$ appearance & \cite{Aguilar:2001ty,KARMEN}($\pi/\mu$ DAR) & $\uparrow$\\    
    $\nu_\mu \rightarrow \nu_e$ appearance & \cite{NOMAD1}($\pi/\mu$ DIF) & 3+1-only \\
    
    $\bar \nu_\mu$ disappearance & \cite{SBMBnubar, CCFR84, MINOSCC2012,MINOSCC2011} ($\pi/\mu$ DIF) & $\downarrow$\\    
    $\nu_\mu$ disappearance & \cite{SBMBnu, CCFR84, CDHS, MINOS2016} ($\pi/\mu$ DIF) & \\ \hline
    $3+1+\mathcal{N}$ & \cite{MiniBooNE:2020pnu} (MiniBooNE BNB $\nu$) &  $\mathcal{N}$  \\ \hline
    \end{tabular}
    \caption{Data sets used in this paper.
    These include reactor, radioactive, decay-at-rest (DAR) and decay-in-flight (DIF) neutrino sources. For a detailed overview of the data sets used in each of the fits see Supplemental~\cref{tab:expts}.}
    \label{tab:fitlist}
\end{table}

In our model, the production and decay of $\mathcal{N}_j$ is due to a dipole portal interaction between left-handed neutrinos, photons, and right-handed HNLs.
The $\mathcal{N}_j$ are added to the SM Lagrangian using the following term~\cite{Masip:2012ke,Magill:2018jla}:
\begin{equation}
\begin{split}
    \mathcal{L} &\supset 
    \mathcal{L}_{SM}  + 
    \sum_{j=1}^{3} \big[ \bar{ \mathcal{N}}_j ( i \slashed{\partial} - M_j ) \mathcal{N}_j \\
    &+ \sum_{\alpha \in \{e,\mu,\tau\}} (d_{\alpha j} \bar \nu_i \sigma_{\mu \nu} F^{\mu \nu} \mathcal{N}_j + h.c.)\big], 
\end{split}
\end{equation}
where the $\nu_i$ correspond to the light neutrino mass states and $F^{\mu \nu}$ is the electromagnetic field strength. The dimension-full $d_{\alpha j}$ couplings control the strength of the electromagnetic interactions between neutrino species, namely the strength of process like $\mathcal{N}_j \to \nu_i \gamma$. Note that $d_{\alpha j}$ reflects the effective dipole coupling of $\mathcal{N}_j$ to the weak eigenstate $\nu_\alpha$.

This leads to two production mechanisms for $\mathcal{N}_j$: coupling to virtual photons produced in meson decays, such as $\pi^0$, and Primakoff upscattering of active neutrinos to $\mathcal{N}$ as they traverse material.
Feynman diagrams for the two production (left, middle) and decay (right) processes are shown in Fig.~\ref{fig:feynman}. 

In our analysis, we considered only production and decay of the third mass state $\mathcal{N}$.
This follows if $d_{\alpha j}$ is the same for all generations and is found to be sufficiently small, such that upscattering is rare, because then the small masses of states 1 and 2 lead to lifetimes that are too long for an SBL experiment to observe decay. In this case, we define $d \equiv d_{\alpha j}$ as a universal coupling.
An alternative explanation if $d$ is found to be large is that the $d_{\alpha j}$ vary with family member, suppressing decays of the first and second mass states. In this case, we define $d \equiv d_{\alpha 3}$, where the coupling of $\mathcal{N}_3$ to all light neutrino species is assumed to be the same.
In the decay, the polarization of $\mathcal{N}$ must be considered~\cite{Formaggio:1998zn,Balantekin:2018azf,deGouvea:2021ual}.
The photon from a right-handed Dirac $\mathcal{N}$ decay has a $(1-\cos\theta)$ distribution, where $\theta$ is the angle between the $\mathcal{N}$ and photon momentum vectors; conversely, a left-handed Dirac $\mathcal{N}$ will decay with a $(1+\cos\theta)$ distribution for the photons.
Production through an unpolarized virtual, off-shell photon yields an equal combination of right-handed and left-handed $\mathcal{N}$, leading to an effectively isotropic photon decay distribution.
For the case of upscattering, where the $\mathcal{N}$ is produced from an interaction with a left-handed muon neutrino, the outgoing $\mathcal{N}$ will be right-handed and the $(1-\cos\theta)$ angular distribution of the photons must be considered.
All three new mass states are related via a mixing matrix to the flavor states.
Calling the new sterile flavors $s_j$, the mass and flavor states are related by:
\begin{equation}
    \nu_\alpha = \sum_i U_{\alpha i} \nu_j + U_{\alpha 3+i} \mathcal{N}_i,~ (\alpha = e, \mu, \tau; s_1, s_2, s_3),
\end{equation}
where $U_{\alpha j}$ is the extended $6\times 6$ neutral-lepton mixing matrix~\cite{Collin:2016aqd}.

\begin{figure}[t]
\begin{center}
\begin{tikzpicture}[line width=1.5 pt, scale=0.5] 
		\node at (-0.8, 1.5) {\large $\pi_0$};
		\draw[fermion] (-1.0,1.) -- (1.5,1);
		\node at (3.95, 2.0) {\large $\nu_\alpha$};		
		\draw[gaugeboson] (1.5,1) -- (3.5,1.0);
		\node at (3.85, 1.0) {\large $\gamma$};		
		\draw[fermionout] (1.5,1) -- (3.5,2.0);
		\node at (3.9, 0.0) {\large $\mathcal{N}$};
		\draw[fermionout] (1.5,1) -- (3.5, 0.0);
		\filldraw (1.5,1) circle (5pt);
\end{tikzpicture}
\begin{tikzpicture}[line width=1.5 pt, scale=0.4]   
		\node at (-0.8, 1.5) {\large $\nu_\alpha$};
		\draw[fermionout] (-1.0,1.0) -- (1.5,1);
		\node at (3.55, 1.5) {\large $\mathcal{N}$};
		\draw[fermionout] (1.5,1) -- (3.5,1.0);
		\draw[gaugeboson] (1.5,1) -- (1.5, -1.8);
		
		\node at (0.9, -0.2) {\large $\gamma$};
		
		\node at (-1.05, -2.0+0.65) {\large $\mathrm{N}$};
		\node at (4.05, -2.0+0.65) {\large $\mathrm{X}$};
		\draw[fermion] (-0.75,-1.8+0.65 ) -- (3.75,-1.8+0.65);
		\draw[fermion] (-0.75,-2.0+0.65 ) -- (3.75, -2.0+0.65);
		\draw[fermion] (-0.75,-2.2 +0.65) -- (3.75,-2.2+0.65);
		\filldraw (1.5,-2.0+0.65) circle (12pt);
\end{tikzpicture}
\begin{tikzpicture}[line width=1.5 pt, scale=0.5]   
		\node at (-0.8, 1.5) {\large $\mathcal{N}$};
		\draw[fermionout] (-1.0,1.) -- (1.5,1);
		\node at (3.55, 0.00) {\large $\gamma$};
		\draw[gaugeboson] (1.5,1) -- (3.5, 0.35);
		\node at (3.55, 1.85) {\large $\nu_\alpha$};		
		\draw[fermionout] (1.5,1) -- (3.5,1.65);
\end{tikzpicture}
\end{center}
\vspace{-0.6cm}
\caption{$\mathcal{N}$ production from $\pi^0$ Dalitz decay (left) and $\nu$ upscattering (middle), and decay (right).}
\label{fig:feynman}
\end{figure}

\section*{Constraints}
Three SBL experiments have relevant limits to $\mathcal{N}$ with mass $>10$ MeV.
While not appearing directly in the fit, the viable solution must fall outside of these limits.
NOMAD and CHARM-II were high-energy neutrino experiments with too small $L/E$ to be sensitive to the 3+1 parameters under discussion.
The NOMAD analysis searched directly for photons from HNL decay~\cite{Gninenko:2012rw}.
CHARM-II could not differentiate electrons from photons, and the limit is derived from comparing $\nu_\mu$-electron elastic scattering (ES) data to the SM prediction~\cite{Coloma:2017ppo}.
At larger $\mathcal{N}$ masses, which are kinematically inaccessible in the former process, contributions from $\nu_\mu$-nucleon upscattering are also present~\cite{Arguelles:2018mtc}.
However, a detailed analysis of this process in CHARM-II has not yet been performed.
LSND has also released $\nu_\mu$-electron scattering results in agreement with the SM, placing a limit on $\mathcal{N}$~\cite{Magill:2018jla}.

Cosmological observations place constraints on additional neutrino species.
In order to alleviate tension with a light sterile neutrino~\cite{Hannestad:2012ky,Lattanzi:2017ubx,Knee:2018rvj,Berryman:2019nvr,Gariazzo:2019gyi,Hagstotz:2020ukm,Adams:2020nue}, one can invoke either noncanonical cosmological scenarios~\cite{Gelmini:2004ah,Hamann:2011ge}
or secret neutrino interactions~\cite{Hannestad:2013ana,Dasgupta:2013zpn,Archidiacono:2014nda,Archidiacono:2015oma,Saviano:2014esa,Chu:2015ipa,Cherry:2016jol,Chu:2018gxk,Song:2018zyl,Farzan:2019yvo,Cline:2019seo}. 
HNL interactions similar to those studied in this model also play an important role in cosmology~\cite{Shakya:2018qzg}, where they may impact Big Bang nucleosynthesis, relax cosmological bounds on neutrino masses~\cite{Escudero:2020ped}, or explain the Hubble tension~\cite{Berbig:2020wve}.

\section*{Oscillation Global Fit}
If the maximum energy of the neutrino source is too small to produce the heavier $\mathcal{N}$ state, then the SBL experiments can observe only 3+1 oscillations.
We refer to this collection of SBL experiments that are not sensitive to $\mathcal{N}$ decay as ``3+1-only" experiments.
The references for relevant ``3+1-only'' experiments used in this analysis are provided in Table~\ref{tab:fitlist} (top), including experiments with anomalies of significance from $2\sigma$ to $4.8\sigma$ and experiments consistent with $\nu$SM oscillations.
The experiments, individually listed, can also be found in Supplemental \cref{tab:expts}. 
We use these experiments to determine the oscillation parameters $\sin^2(2\theta)$ and $\Delta m^2$ in a 3+1-only fit. Notably, the ``3+1-only'' experiments exclude all MiniBooNE results, as we will then use MiniBooNE to fit for the $\mathcal{N}$ parameters $m_\mathcal{N}$ and $d$. 

As shown in Table~\ref{tab:fitlist} (bottom), we used MiniBooNE data to fit for the $\mathcal{N}$ parameters $m_\mathcal{N}$ and $d$, given the oscillation parameters from the 3+1-only fit. MiniBooNE has excesses in three appearance data subsets: neutrino-mode~\cite{MBnu}, antineutrino mode~\cite{MBnubar}, and with an off-axis beam~\cite{MBNumi}.
All three cases are compatible with either 3+1 or HNL explanations.
However, the latter two running modes had low statistics and more limited data releases, so we restricted our fit to the neutrino-mode sample.

For the 3+1-only fit, mixing between heavy neutrinos and the three lightest mass states is constrained to be small by terrestrial measurements at accelerators~\cite{T2K:2019jwa}. Further, oscillations involving the two largest mass states do not contribute to the explaining the anomalies considered in this work.Therefore, we have explicitly assumed no mixing between the two largest mass states and the active states.
The only relevant squared-mass-splitting $\Delta m^2$ is between the lightest mostly sterile and the mostly active states, where the masses of the latter are assumed to be degenerate and negligible.
We concentrated on the $4\times4$ neutral-lepton-mixing submatrix that relates the lightest mass states to their flavor states.
For $U_{\alpha k}$, where $\alpha$ is the flavor and $k$ is the mass state, the mixing angles for the appearance and disappearance oscillation signatures are not independent: $\sin^22\theta_{ee} = 4(1-|U_{e4}|^2)|U_{e4}|^2$ (electron flavor disappearance); $\sin^22\theta_{\mu \mu } = 4(1-|U_{\mu 4}|^2)|U_{\mu 4}|^2$ (muon flavor disappearance); and $\sin^2 2\theta_{\mu e} = 4|U_{\mu4}|^2 |U_{e4}|^2$ (appearance).
This implies that the electron and muon flavor disappearance signals must be consistent with the $\nu_\mu \rightarrow \nu_e$ appearance signal, limiting the range of $\sin^2 2\theta_{e\mu}$.

This analysis employed the 3+1 global fitting code described in Ref.~\cite{Diaz:2019fwt}.
The list of experiments used in the 3+1-only fit can be found in Supplemental \cref{tab:expts}.
Compared to Ref.~\cite{Diaz:2019fwt}, we have added new disappearance results from the STEREO experiment~\cite{PhysRevD.102.052002} and updated PROSPECT results~\cite{PhysRevD.103.032001}. 
In this update, we have not included the results from NEUTRINO-4~\cite{Serebrov:2018vdw}, since the collaboration has not provided an appropriate data release.
We have also not included the latest result from IceCube~\cite{Aartsen:2020iky,Aartsen:2020fwb}, which shows a preferred region at 90\% CL compatible with our light sterile neutrino best-fit point, since the collaboration has not provided enough information to reproduce the analysis.
To reiterate, the 3+1-only global fit omitted the MiniBooNE neutrino-mode, antineutrino-mode, and off-axis appearance data sets.

The best-fit parameters are $\Delta m^2 = 1.32\,{\rm eV}^2$ and $\sin^2 2\theta_{e\mu} = 6.9 \times 10^{-4}$.
In past 3+1 fits, the tension between the appearance and disappearance data sets~\cite{Diaz:2019fwt}, as measured using the PG test, has been very high, with a probability of $4\times 10^{-6}$ (4.5$\sigma$) that the data are explained by the same parameters. 
Without MiniBooNE appearance in the fit, the probability increases to $7 \times 10^{-3}$ ($2.5\sigma$). 
Thus, the tension is, in large part, due to the MiniBooNE appearance data set, which we hypothesize has the additional component of $\mathcal{N}$-decay, and, hence, poor agreement with 3+1-only.

\section*{MiniBooNE Analysis}
In order to fit MiniBooNE data for $\mathcal{N}$ decay, we wrote a Monte Carlo simulation for the production and decay of $\mathcal{N}$ in the Booster Neutrino Beam in neutrino mode.
Two processes were included for production:  Dalitz-like $\pi^0$ decay and Primakoff upscattering $\nu A \to \mathcal{N} A$.
The latter is by far the dominant $\mathcal{N}$-production mode for $10\,{\rm MeV}< m_\mathcal{N}<1000\,{\rm MeV}$.
Therefore, we neglected the $\pi^0$ decay contribution throughout this study.
For the Primakoff mode, we generated incident $\nu_\mu$ and $\nu_e$ events from the MiniBooNE neutrino-mode flux~\cite{Aguilar_Arevalo_2009}.
We then simulated the upscattering rate on both standard upper-continental crust nuclei~\cite{EarthCrust} and the MiniBooNE CH$_2$ detector medium, using Eq.~A6 in Ref.~\cite{Magill:2018jla} to calculate the total interaction rate and momentum transfer.
This process produced a sample of right-hand-polarized $\mathcal{N}$ events, predominately forward peaked due to the $1/t^2$ dependence of the differential cross section. 

\begin{figure}[h]
\begin{center}
\includegraphics[width=\columnwidth]{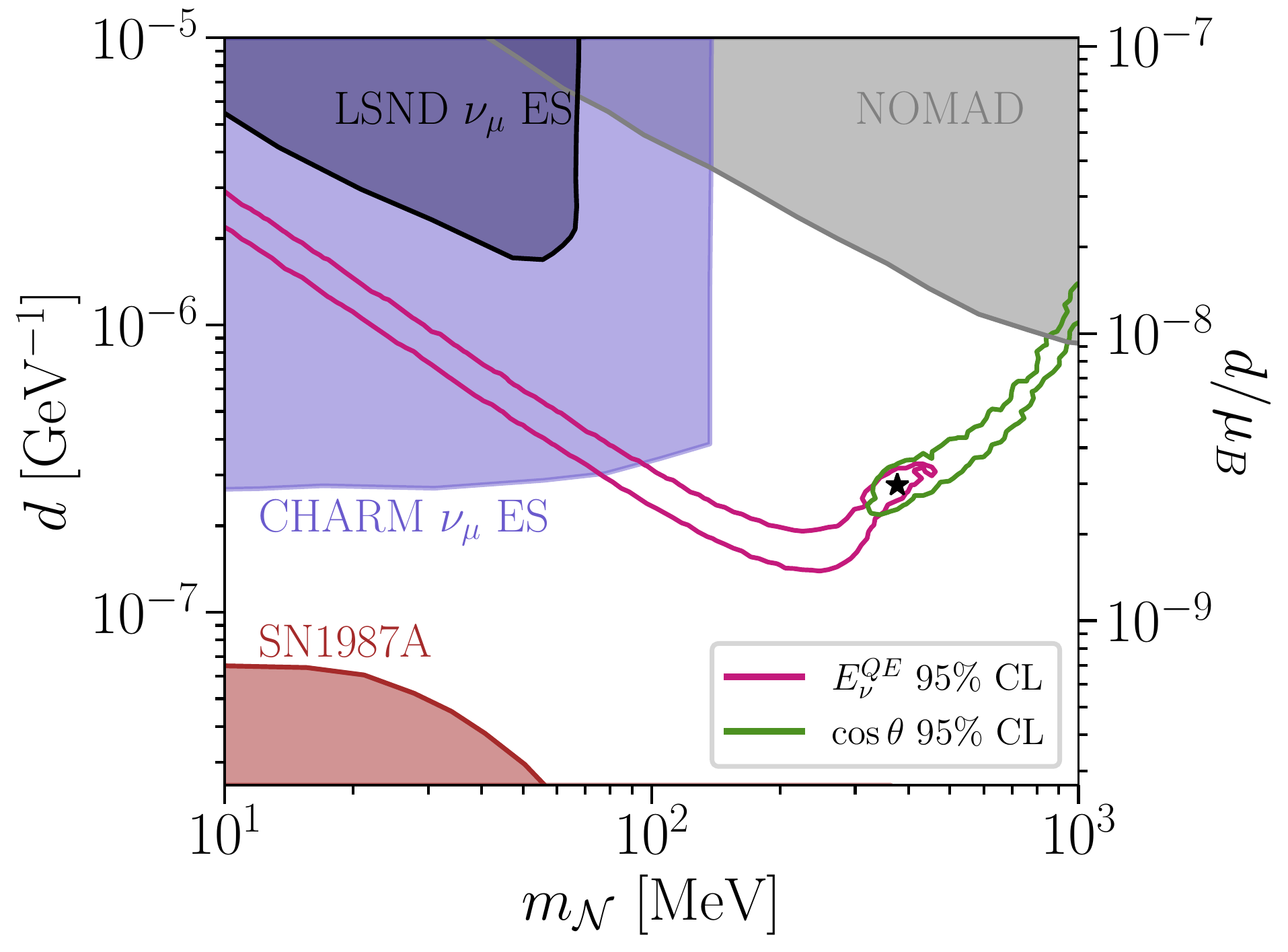}
\end{center}
\vspace{-0.6cm}
\caption{Preferred regions to explain the MiniBooNE excess in $E_\nu^{QE}$ (pink) and $\cos \theta$ (green) as a function of dipole coupling and $\mathcal{N}$ mass. The black star indicates $\{d,~m_\mathcal{N}\} = \{2.8 \times 10^{-7}\;\text{GeV}^{-1},~376\;\text{MeV}\}$, which lies in the joint 95\% CL allowed region for both distributions. Constraints from other experiments are also shown at the 95\% CL. }
\label{fig:couplingvsMass}
\end{figure}

Simulated $\mathcal{N}$ which enter the MiniBooNE detector were forced to decay into a photon and a neutrino, taking into account polarization, and weighted by the decay probability.
To incorporate the detector efficiency, $eff$, we performed a linear fit to the reconstructed gamma-ray efficiency as a function of true energy~\cite{MiniBooNE},  $eff=(-0.12\,{\rm GeV^{-1}})* E_{true} + 0.29$, which we used to weight the $\mathcal{N} \to \nu \gamma$ events.
The true energy and angle of the photons were smeared independently according to the resolution given by the MiniBooNE collaboration. 
More details on the simulation can be found in Appendix~\ref{app:simdetails}.

Ideally one would fit the background-subtracted 2-D distribution of visible energy, $E_{v}$, vs. scattering angle, $\theta$, of the MiniBooNE events to the prediction from $3+1+\mathcal{N}$-decay.
However, the systematic errors for this distribution have not been released by MiniBooNE.
They are only available for $E_\nu^{QE}$, which is a combination of $E_{v}$ and $\theta$ given by~\cite{PhysRevD.103.052002}
\begin{equation}
    E_\nu^{QE}=\dfrac{2(M_{n}-B) E_{v}-((M_n-B)^{2}+M_{e}^{2}-M_{p}^{2})}{2((M_{n}-B)-E_{v}+\sqrt{(E_{v}^{2}-M_{e}^{2})} \cos\theta)},
    \label{enuqe}
\end{equation}
where $M_n$, $M_p$, and $M_e$ are the neutron, proton and electron masses, and $B$ is the binding energy of carbon.
This formula accurately describes the neutrino energy in the case of two-body charged-current neutrino scattering with no final state interactions, assuming the neutrinos enter the detector along the axis from which $\theta$ is measured.
Thus, it is applicable to the oscillation component of the excess.
Though $E_\nu^{QE}$ has no physical meaning when applied to the photons from $\mathcal{N}$ decay, the decay kinematics cause most events to be show up at low $E_\nu^{QE}$.
We performed a fit to the MiniBooNE excess in $E_\nu^{QE}$ using statistical and systematic uncertainties.    
We also performed a separate fit to the scattering angle distribution, although only statistical uncertainty is available in this case.

To isolate the decay component, we subtracted from the MiniBooNE excess the predicted contribution of the oscillation component, which was determined from the $3+1$-only global fit without MiniBooNE data.
The remaining excess was fit to the model for dipole production, decay, and observation in the detector as described above.
Fig.~\ref{fig:couplingvsMass} shows confidence regions for both fits in $\{d,m_\mathcal{N}\}$ parameter space, computed assuming Wilks' theorem with two degrees of freedom is valid for the test statistic $\chi^2(d,m) - \min_{d,m}(\chi^2(d,m))$~\cite{WilksThm}.
We found a region of parameter space consistent with both distributions at the 95\% CL near $d=3 \times 10^{-7}$ GeV$^{-1}$ and $m_\mathcal{N} = 400$ MeV.  

\section*{Results}
Fig.~\ref{fig:couplingvsMass} shows that the allowed regions from MiniBooNE fits are substantially lower in $d$ than the NOMAD or LSND limits.
The overlapping solution is also at substantially higher $m_\mathcal{N}$ than kinematically accessible by LSND.
Supernova results~\cite{Magill:2018jla} set limits in a band from approximately $d=10^{-7}$ to $10^{-11}$~GeV$^{-1}$, which is below the solution we found for MiniBooNE.
Thus, our preferred region  lies in a window of allowed parameters.  

We now consider an example HNL decay contribution for $d = 2.8 \times 10^{-7}$ GeV$^{-1}$ and $m_\mathcal{N}$ = 376~MeV, indicated by the star in Fig.~\ref{fig:couplingvsMass}.
This corresponds to the best fit to the $E_\nu^{QE}$ distribution within the joint 95\% CL allowed region from the $E_\nu^{QE}$ and $\cos \theta$ fits.
Table~\ref{tab:fitchisq} shows the $\chi^2$ values for the $3+1$ and $3+1+\mathcal{N}$-decay fits to both distributions, indicating significant improvement for the $3+1+\mathcal{N}$-decay model.
The global oscillation fit gives tight constraints requiring $\Delta m^2 \approx 1.32$ eV$^2$, but allows values of $\sin^2 2 \theta_{\mu e} \in [3\times 10^{-4},2 \times 10^{-3}]$ at the 90\% CL.
The same $\mathcal{N}$ decay fit procedure outlined above has been performed for each end of the allowed $\sin^2 2 \theta_{\mu e}$ range.
In each case we again examined the $\{d,m_\mathcal{N}\}$ point that best fits the $E_\nu^{QE}$ distribution within the joint 95\% CL region.
The $\chi^2$ values for these fits are also given in Table~\ref{tab:fitchisq}, indicating a preference for a smaller oscillation contribution in MiniBooNE in order to explain both the $\cos \theta$ and $E_\nu^{QE}$ distributions via $\mathcal{N}\to \nu \gamma$. 
Table~\ref{tab:fitchisq} also gives the $\chi^2$ values for the null case, with neither eV-scale oscillations nor HNL decay.

\begin{table}[ht]
    \centering
    \begin{tabular}{|c|c|c|c|c|}
    \hline
  \multicolumn{1}{|c|}{Parameters} &  \multicolumn{4}{|c|}{$\chi^2/dof$}\\ \hline  
    ($\sin^2 2\theta$,$d$,$m_\mathcal{N}$)& \multicolumn{2}{|c|}{$3+1+\mathcal{N}$} & \multicolumn{2}{|c|}{$3+1$}\\
    \cline{2-5}
    & ${E_\nu^{QE}}$  &  ${\cos \theta}$ &  ${E_\nu^{QE}}$ & ${\cos \theta}$\\
    \hline
     (0.30, 3.1, 376) & 5.7/8 & 32.1/18 & 30.5/10 & 86.4/20 \\
     \hline
     (0.69, 2.8, 376) & 7.9/8 & 31.4/18 & 27.3/10 & 71.8/20 \\
     \hline
     (2.00, 5.6, 35) & 20.2/8 & 36.7/18 & 27.6/10 & 40.8/20  \\
     \hline
     (0, 0, 0) & 34.1/10 & 99.4/20 & same & same  \\
     \hline
    \end{tabular}

    \caption{$\chi^2$/dof values for $3+1$ and $3+1+\mathcal{N}$-decay models obtained by comparing expectations to the MiniBooNE excess in $E_\nu^{QE}$ and $\cos \theta$. 
    The parameters in column one refer to ($\sin^2 2\theta_{\mu e}\times 10^{-3}$, $d\times 10^{-7}$ [GeV$^{-1}]$, $m_\mathcal{N}$ [MeV]).
    The mass splitting is 1.32 eV$^2$ in all cases. 
    The null case (no oscillations and no HNL decay) is also shown in the last row.}
    \label{tab:fitchisq}
\end{table}

\begin{figure}[tb]
\begin{center}
\includegraphics[width=0.5\columnwidth]{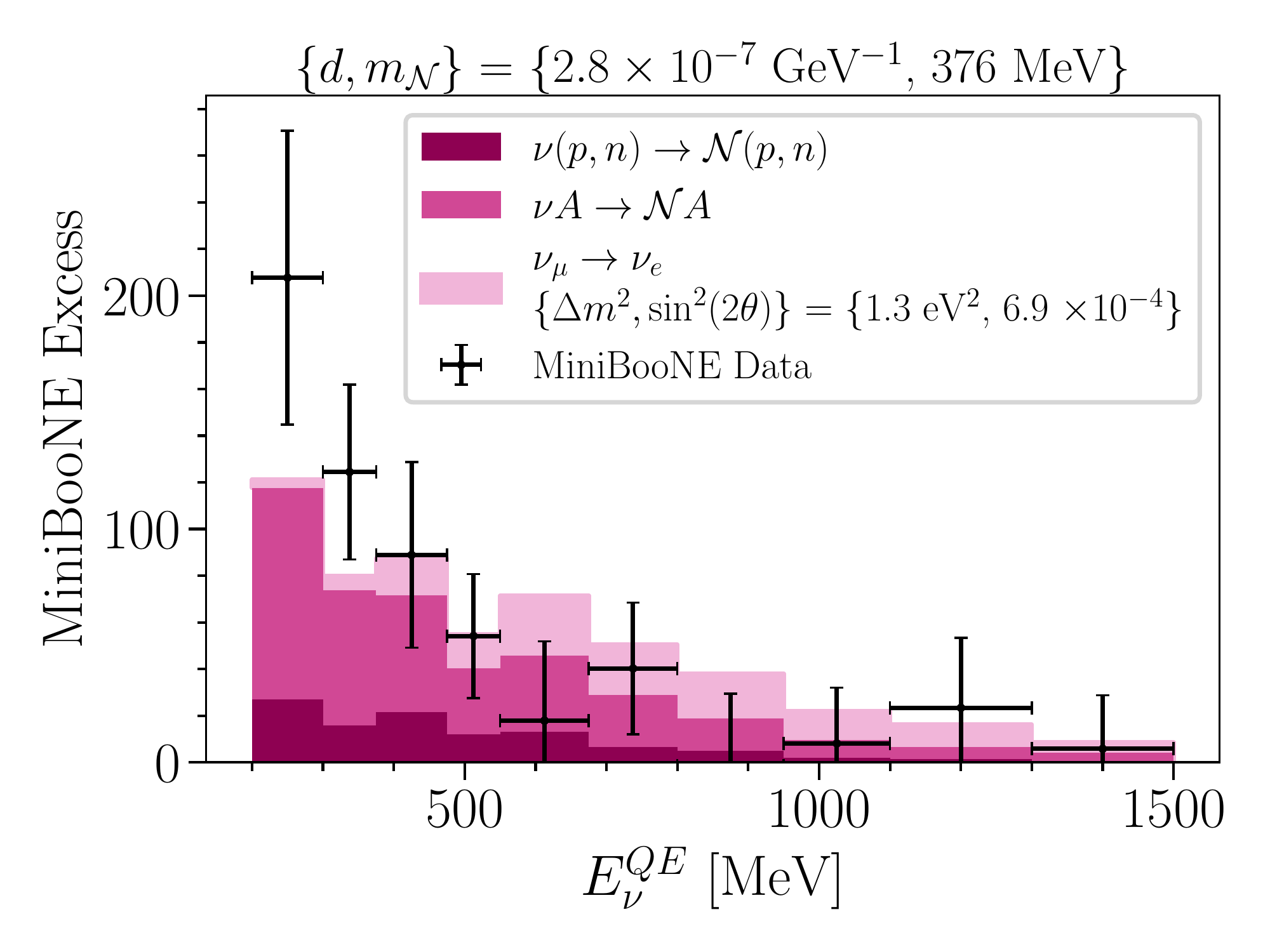}
\includegraphics[width=0.5\columnwidth]{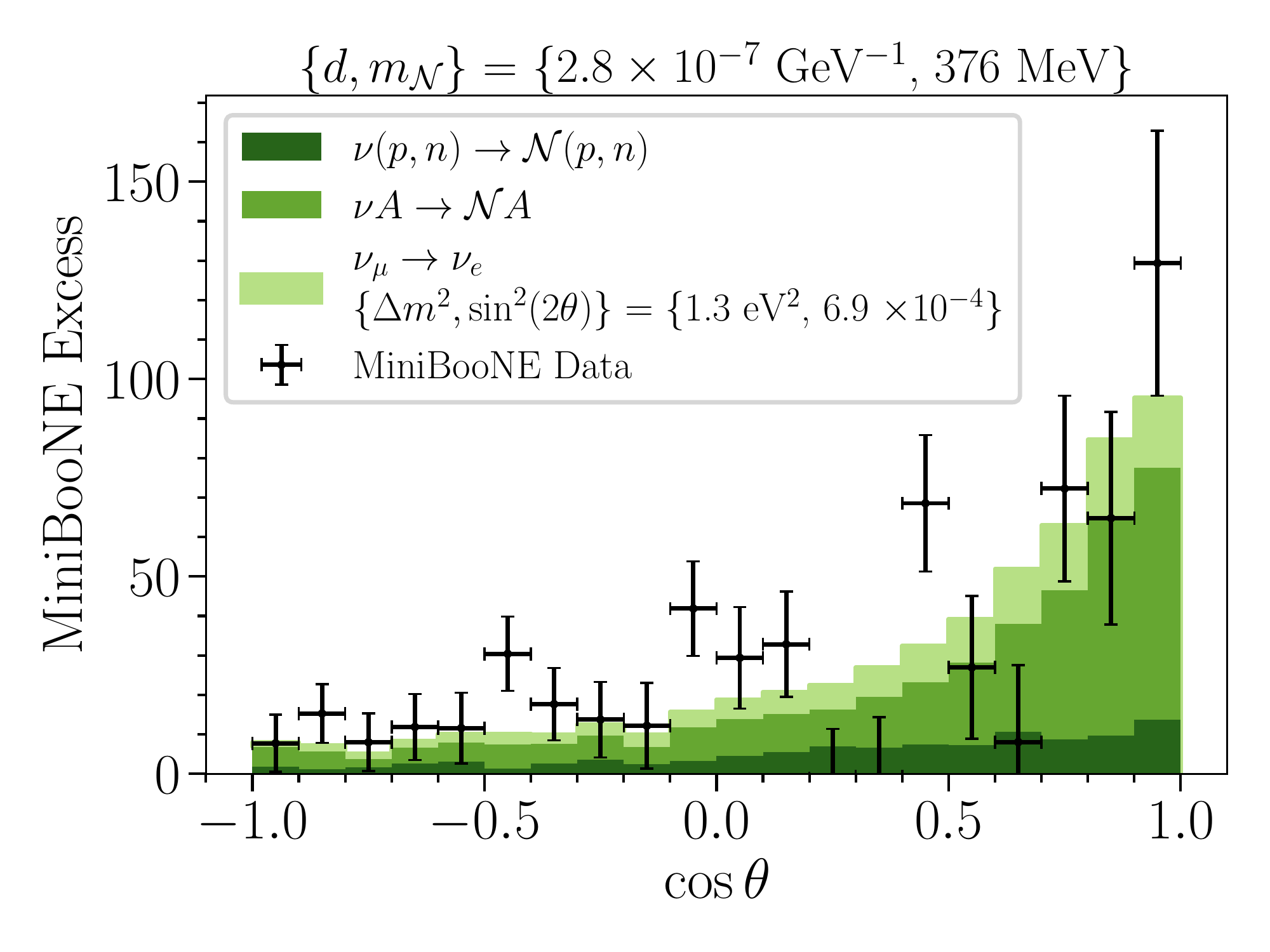}
\end{center}
\vspace{-0.6cm}
\caption{ $E_\nu^{QE}$ (left) and $\cos \theta$ (right) distributions of the MiniBooNE excess for a representative point of the $3+1+\cal{N}$-decay model.
The error bars on the energy distribution include systematic and statistical errors, while for the angular distribution only statistical errors are included.
\label{fig:energyangle}}
\end{figure}

Fig.~\ref{fig:energyangle} presents the MiniBooNE excess in $E_\nu^{QE}$ (left) and $\cos \theta$ (right) compared with the model prediction. 
This figure includes both the global fit oscillation contribution for $\Delta m^2 = 1.3$~eV$^2$ and $\sin^2 2 \theta_{ee} = 6.9 \times 10^{-7}$, and HNL decay contribution for $d = 2.8 \times 10^{-7}$~GeV$^{-1}$ and $m_\mathcal{N}$ = 376~MeV. 
We re-emphasize here that the oscillation contribution shown on these plots comes from a global fit to the 3+1-only model not including MiniBooNE data.
Good agreement is observed for both distributions, especially noting again the lack of systematic errors for the angular distribution, which dominate over statistical errors in the MiniBooNE electron-like analysis~\cite{MiniBooNE:2020pnu}. 

Along with energy and angle, it is essential for the model to be consistent with the recently published timing distribution of the MiniBooNE excess with respect to the proton-beam bunch.
The excess events occur within $\pm 4$~ns of the observed $\nu_\mu$ events.
The time of flight depends on the location at which the HNL is produced. For the preferred $m_\mathcal{N} \sim 400\,$MeV, the majority of $\mathcal{N}$ events come from upscattering within the MiniBooNE detector followed by $\mathcal{N}$ decay after travel distances of $\mathcal{O}(50\,{\rm cm})$. This leads to timing delays of $< 2\,{\rm ns}$, well within the MiniBooNE constraint. 

\section*{Conclusion}
We have explored a 3+1+$\mathcal{N}$-decay model through fits to the 3+1-only and MiniBooNE-neutrino-mode data sets.
The former yields best-fit oscillation parameters of  $\Delta m^2 = 1.3$ eV$^2$ and 
$\sin^2 2\theta_{e\mu}= 6.9 \times 10^{-4}$.
The latter narrows the HNL mass to be within $300-400$~MeV and the dipole coupling strength to be within $2.5-3.5 \times 10^{-7}\,\rm{GeV}^{-1}$.
This model produces a $\Delta \chi^2$/dof improvement of 19.3/2 (40.3/2) compared to the global 3+1 scenario for the fit to the MiniBooNE energy (angular) distribution.      
This large improvement in $\Delta \chi^2$ motivates a more detailed analysis by MiniBooNE.
Ideally, the experiment would perform a joint fit to the two-dimensional visible energy and angle distribution, using a full covariance matrix.

Our model also makes very specific predictions for the experiments now running in Fermilab's Short-Baseline Neutrino Program~\cite{SBNProp}.
These experiments make use of liquid-argon time-projection chambers (LArTPCs) that can separate photon showers from electron showers with $\sim 85\%$ accuracy~\cite{Adams_2020,uB_egamma}.
Because the experiments run in the same beamline and are located within $\sim$50~m of MiniBooNE, the flux is nearly identical.
Thus, the ratio of oscillation to HNL decay contributions for the far detectors --- MicroBooNE and ICARUS --- will be very similar to that of the MiniBooNE case presented here, with {\bf $\sim 75\%$} of the excess events predicted to be single photons.
The photon signature will have large backgrounds even in an LArTPC detector, especially from neutral current $\Delta$ baryon production with decays to one or two photons plus a neutron, as well as photons from neutrino interactions produced outside the detector~\cite{uB_glee}.
However, this background rate will be well constrained from reconstructed $\Delta$-decay events with a proton.
Also, the energy-angle correlation of the photon in the decay, which depends strongly on $m_{\mathcal{N}}$, can be used for background rejection since this parameter is predicted to have a narrow range of values.
The oscillation rates for MicroBooNE and ICARUS are predicted to be low.
Within statistics, we predict that no clear oscillation signature will be observed. 

Beyond the SBN program, our model can also produce signatures in $\nu_\mu$ ES searches at MINER$\nu$A and NO$\nu$A, as well as dedicated searches for single photons in T2K~\cite{Abe:2019cer}.
Additionally, at neutrino energies of  $\mathcal{O}(100\,\rm{GeV})$ the long decay length produces ``double-bang'' morphologies in IceCube and CCFR~\cite{Coloma:2017ppo,Coloma:2019qqj,CCFR:1992vtp,CCFR:1992ajg}, which would be a smoking gun signature for our model.

In summary, we have presented a new physics model including neutrino-partners with masses of $\mathcal{O}$(1~eV) that participate in oscillations and $\mathcal{O}$(100~MeV) that decay to single photons.
This model can simultaneously explain the MiniBooNE anomaly and relieve tension in the global experimental picture for 3+1 oscillations.
The results indicate very narrow ranges of HNL decay and oscillation parameters; thus, this is a highly predictive result that can be further tested by existing experiments in the near future.

\section*{Acknowledgements}

MHS is supported by NSF grant PHY-1707971. NSF grant PHY-1801996 supported 
CAA, JMC, AD and NWK for this work.  Additionally,
CAA is supported by the Faculty of Arts and Sciences of Harvard University and NWK is supported by the NSF Graduate Research Fellowship under Grant No. 1745302. MAU is supported by the Department of Physics at the University of Cambridge and SV is supported by the STFC. We thank the MiniBooNE Collaboration for useful input, and Gabriel Collin and William Louis for comments on the draft of this paper.
Finally, we thank Matheus Hostert and Ryan Plestid for useful discussions.

\bibliographystyle{apsrev}
\bibliography{neutrissimo}

\pagebreak
\clearpage


\onecolumngrid
\appendix

\ifx \standalonesupplemental\undefined
\setcounter{page}{1}
\setcounter{figure}{0}
\setcounter{table}{0}
\setcounter{equation}{0}
\fi

\renewcommand{\thepage}{Supplemental Material-- S\arabic{page}}
\renewcommand{\figurename}{SUPPL. FIG.}
\renewcommand{\tablename}{SUPPL. TABLE}

\renewcommand{\theequation}{A\arabic{equation}}
\clearpage

\begin{center}
\textbf{\large Supplemental Material}
\end{center}

\section{Further Explanation of the MiniBooNE $\mathcal{N}$ Simulation \label{app:simdetails}}

\subsection*{Primakoff Upscattering Simulation}

In order to simulate the Primakoff process $\nu A \to \mathcal{N} A$ for a given dipole $d$ coupling and HNL mass $m_\mathcal{N}$, we first generated a sample of $5 \times 10^5$ $\nu_\mu$ and $\nu_e$ events with energies according to the MiniBooNE neutrino-mode flux from~\cite{Aguilar_Arevalo_2009} and initial azimuthal angles generated randomly in $\cos \theta$ within a cone encompassing the MiniBooNE detector.
We chose a scattering location uniformly in column density. For the purposes of this study, we have taken the detector to be a sphere of CH$_2$ with a radius of 6.1 m, surrounded by a concentric sphere of air with radius 9.1 m, intended to represent the detector hall.
The rest of the volume is taken to be standard upper-continental crust. We next select a nucleus for the scattering event according to atomic abundances and scattering cross sections. 
If the scattering event took place in the dirt, we use atomic abundances from~\cite{EarthCrust}, which are reproduced in Table~\ref{tab:crustnuclei}.
Upscattering events inside the MiniBooNE detector happen exclusively off of CH$_2$. 
Events almost never happen in the air surrounding MiniBooNE due to the low density.
Scattering cross sections are calculated by numerically integrating the differential cross section, adapted from Ref.~\cite{Brdar:2020quo}:

\
\begin{equation}\label{eq:prima_dsigdt}
\begin{split}
\frac{d \sigma}{d t} = \frac{2 \alpha d^2 }{m} \bigg[ & F_1^2(t) \Big( \frac{1}{E_r} - \frac{1}{E_\nu} + m_\mathcal{N}^2 \frac{E_r - 2 E_\nu - M}{4 E_\nu^2 E_r M} + m_\mathcal{N}^4 \frac{E_r - M}{8 E_\nu^2 E_r^2 M^2} \Big) \\
+ &\frac{F_2^2(t)}{4 M^2} \Big( \frac{2 M }{E_\nu^2} ( (2 E_\nu - E_r)^2 - 2 E_r M ) + m_\mathcal{N}^2 \frac{E_r - 4 E_\nu}{E_\nu^2} + \frac{m_\mathcal{N}^4}{E_\nu^2 E_r}\Big)\bigg],
\end{split}
\end{equation}
where $\alpha$ is the fine structure constant, $d$ is the dipole coupling, $E_\nu$ is the SM neutrino energy, $m_\mathcal{N}$ is the mass of the heavy neutrino, $M$ is the target mass, $t = -(p_\mathcal{N} - p_\nu)^2$ is the momentum transfer, $E_r = -t/2M$ is the target recoil energy, and $F_{1/2}(t)$ are the charge/magnetic target form factors, respectively.
Note that the term proportional to $E_r m_N^4$ in the $F_1$ line only exists for spinless nuclei, and must be replaced for nonzero spin nuclei~\cite{Masip:2012ke}.
In the case of coherent scattering off of a nucleus, $F_1$ receives a $Z^2$ enhancement and is therefore dominant over $F_2$; which has therefore been neglected for this study.
Here $F_1$ is given by the dipole approximation
\begin{equation}
F_1^A = \frac{1}{(1 + (r_A^2 q^2)/12)^2},
\end{equation}
where $r_A$ is the nuclear radius of the target. 
In the case of inelastic scattering off of nucleons, the form factors are calculated by solving the following system of equations, repeated here from Appendix A of Ref.~\cite{Magill:2018jla}. 
\begin{equation}
\begin{split}
&G^{\{p,n\}}_E = F_1^{\{p,n\}} - \frac{Q^2}{4 m_{\{p,n\}}^2} F_2^{\{p,n\}} = \{G_D,0\}, \\
&G^{\{p,n\}}_E = F_1^{\{p,n\}} - \frac{Q^2}{4 M^2} F_2^{\{p,n\}} = \mu_{\{p,n\}} G_D, \\
&G_D = \frac{1}{(1 + Q^2/0.71~\text{GeV}^2)^2}, \\
& \mu_{\{p,n\}} = \{2.793,-1.913\}.
\end{split}
\end{equation}
The total cross section for scattering off of a nuclear target $A$ is given by the incoherent sum of the nuclear and nucleon scattering cross sections, namely
\begin{equation}
\sigma_{A} = Z^2 \int_{t_{min}}^{t_{max}} \frac{d \sigma_{\nu A \to \mathcal{N} A}}{dt} dt + Z\int_{t_{min}}^{t_{max}} \frac{d \sigma_{\nu p \to \mathcal{N} p}}{dt} dt + (A-Z) \int_{t_{min}}^{t_{max}} \frac{d \sigma_{\nu n \to \mathcal{N} n}}{dt} dt,
\end{equation}
where the lower bounds for each integral are given in Appendix C of Ref.~\cite{Magill:2018jla} and the upper bounds are calculated by requiring physical scattering angles $|\cos(\theta)| < 1$.
From this we calculate the probability of scattering off a given nucleus $A_k$ (with atomic fractional abundance in dirt/air/CH$_2$ $F_k$) as
\begin{equation}
P_k = \frac{F_k \sigma_{A_k}}{\sum_{k'} F_{k'} \sigma_{A_{k'}}}.
\end{equation}

\begin{table}[b]
    \centering
    \begin{tabular}{|c|c|c|c|c|c|}
        \hline
        Nucleus & Z & A & Crust Mass Fraction & Crust Atomic Fraction & Nuclear Radius [MeV$^{-1}$] \\  
        \hline
        O & 8 & 16 & 0.466 & 0.627 & 0.00218 \\
        Si & 14 & 28 & 0.277 & 0.213 & 0.00252 \\
        Al & 13 & 27 & 0.081 & 0.065 & 0.00247 \\
        Fe & 26 & 56 & 0.05 & 0.019 & 0.00301 \\
        Ca & 20 & 40 & 0.037 & 0.02 & 0.00281 \\
        K & 19 & 39 & 0.027 & 0.015 & 0.00277 \\
        Na & 11 & 23 & 0.026 & 0.024 & 0.00241 \\
        Mg & 12 & 24 & 0.015 & 0.013 & 0.00247 \\
        Ti & 22 & 48 & 0.004 & 0.002 & 0.0029 \\
        P & 15 & 31 & 0.001 & 0.001 & 0.00257 \\
        \hline
    \end{tabular}
    \caption{Relevant parameters of the ten most abundant nuclei in the Earth's upper crust according to \cite{EarthCrust}}
    \label{tab:crustnuclei}
\end{table}

At this stage, we decide whether each upscattering event occurred coherently off of the nucleus or inelastically off of a proton or neutron by considering the relative cross sections.
We then pull a random momentum transfer from Eq.~\ref{eq:prima_dsigdt}.
Once we have chosen a value for $t$, the heavy neutrino energy and scattering angle are fixed~\cite{Masip:2012ke}:
\begin{gather}
E_\mathcal{N} = E_\nu - E_r \\
\cos(\theta) = \frac{E_\nu - E_r - M E_r / E_\nu - m_\mathcal{N}^2/2E_\nu}{\sqrt{E_\nu^2 + E_r^2 - 2E_\nu E_r - m_\mathcal{N}^2}}.
\end{gather}
The $1/t^2$ dependence of Eq.~\ref{eq:prima_dsigdt} creates a preference for $E_\mathcal{N} \approx E_\nu$ and $\cos(\theta) \approx 1$.
At this stage, if the scattering angle of the $\mathcal{N}$ is greater than the solid angle of the MiniBooNE detector (considering the scattering location), the event is rejected.
If not, we multiply the existing weights by the probability that the heavy neutrino decays via $\mathcal{N} \to \nu \gamma$ in MiniBooNE, which is given by the following expression~\cite{Magill:2018jla}:
\begin{gather} \label{eq:decaylength}
P_{decay}=\exp(\frac{-L_{enter}}{L_{decay}})-\exp(\frac{-L_{exit}}{L_{decay}}) \\
L_{decay} = 4 \pi \hbar c \frac{\beta E_\mathcal{N}}{d^2 (m_\mathcal{N}c)^4},
\end{gather}
where $L_{enter/exit}$ denote the distance between the creation point of the HNL and the entry/exit point in the MiniBooNE detector, respectively.
The culmination of this simulation chain is a weighted sample of $\mathcal{N}$ events which come from the Primakoff upscattering process and decay in the MiniBooNE detector.
The last remaining step is to calculate the POT required to get $N = 5 \times 10^5$ upscattering events along the BNB beamline.
This will depend on the dipole coupling and heavy neutrino mass in general, and can be calculated using
\begin{equation}
N = \text{POT} \int_{E_\nu = 0~\text{GeV}}^{10~\text{GeV}} d E_\nu  \phi(E_\nu)  \sum_{\text{targets}~A} \sigma_A(E_\nu) n_A,
\end{equation}
where $\phi(E_\nu)~[ \nu / \text{POT}]$ is again the neutrino flux in MiniBooNE~\cite{Aguilar_Arevalo_2009}, $\sigma_A~[\text{cm}^2]$ is the total upscattering cross section for nuclear target $A$, and $n_A [\text{nuclei}/\text{cm}^2 ]$ is the column density for $A$ along the beamline.

\begin{figure}[h]
\begin{center}
\includegraphics[width=0.4\columnwidth]{figures/FinalPlotsNick/ConfidenceRegions0.pdf}
\includegraphics[width=0.4\columnwidth]{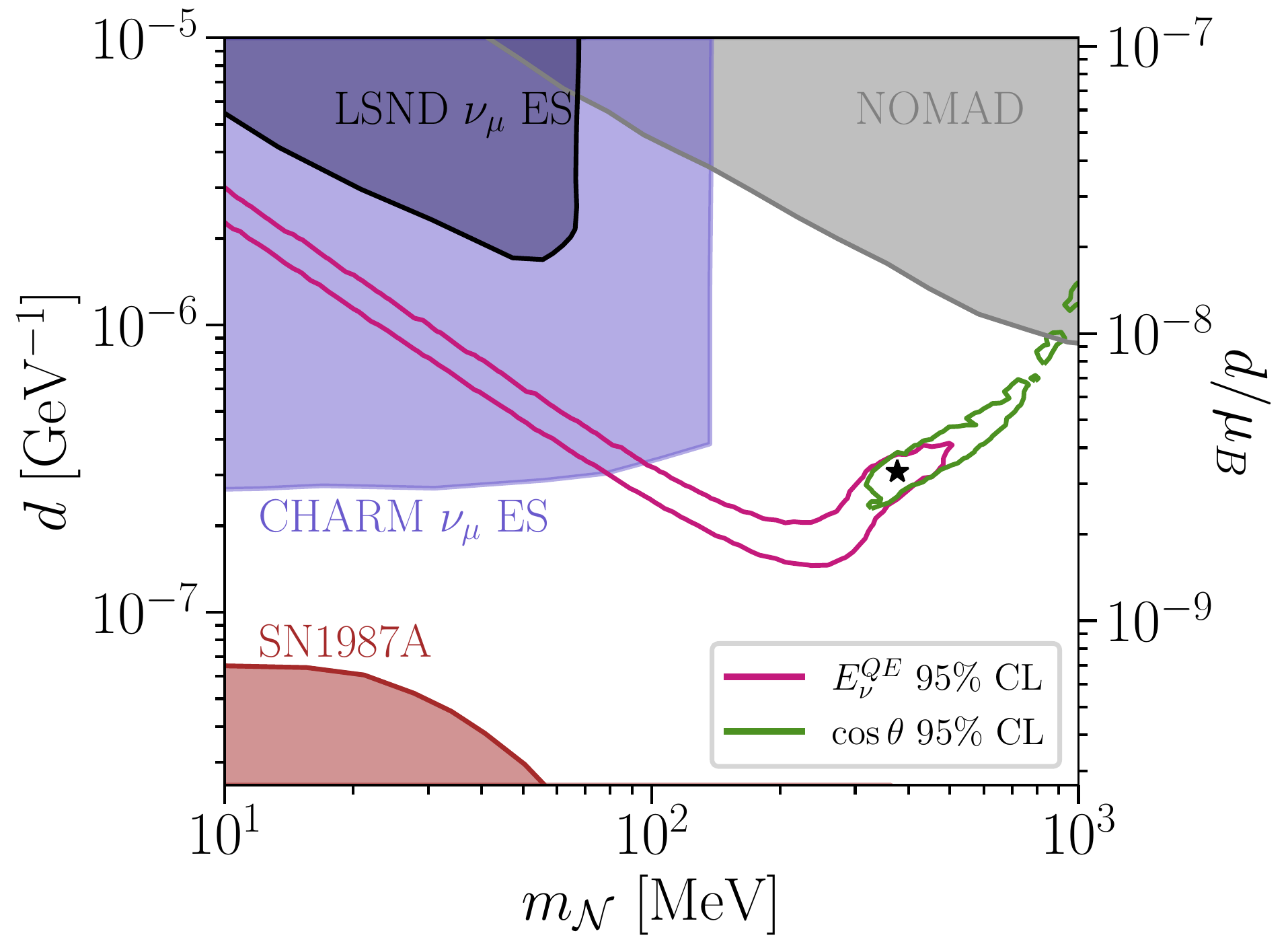}
\includegraphics[width=0.4\columnwidth]{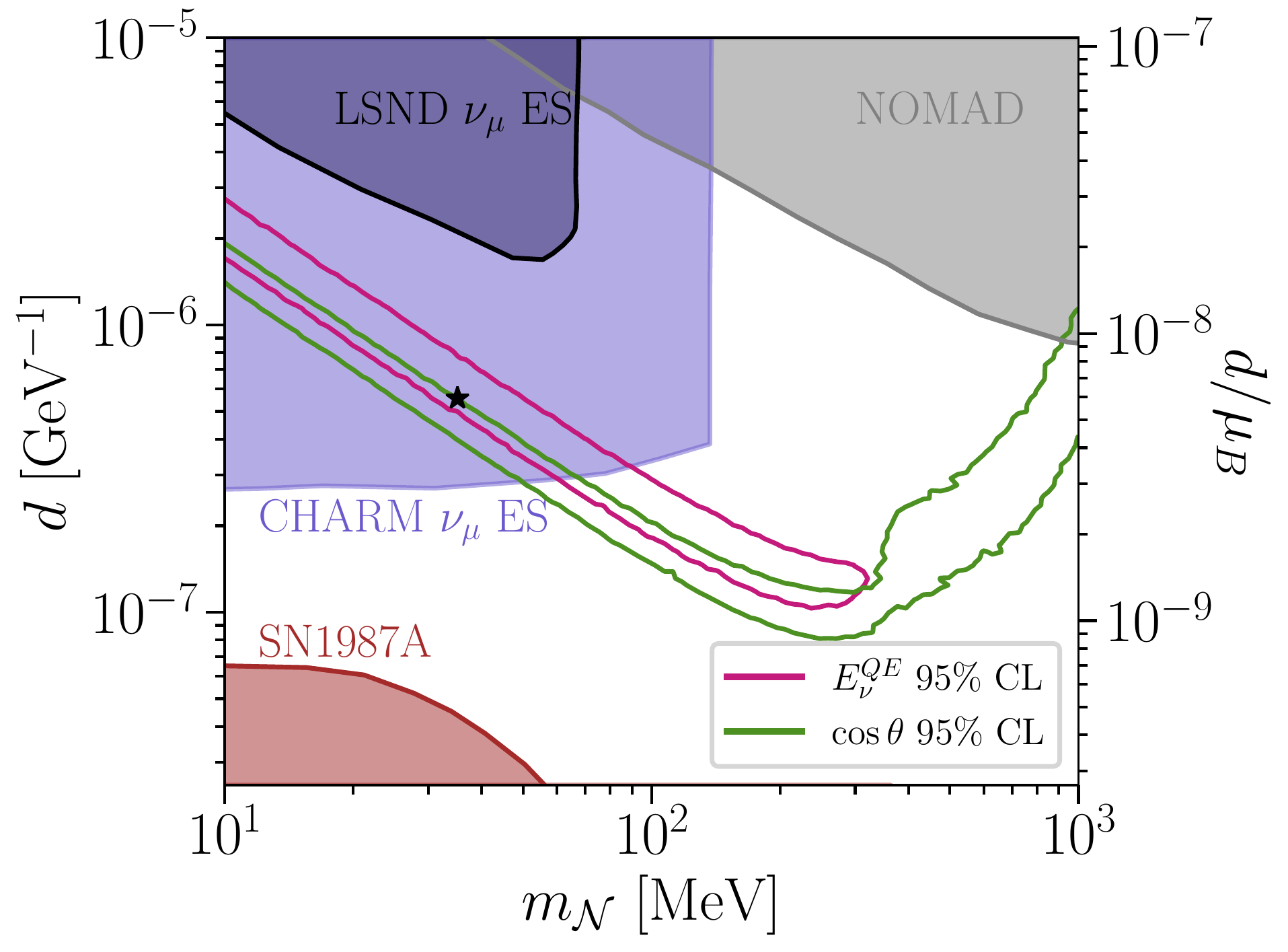}
\end{center}
\caption{ 95\% CL allowed regions in HNL dipole model parameter space from energy and angular fits to the MiniBooNE excess, considering an oscillation contribution with $\Delta m^2 = 1.3$ eV$^2$ and three different amplitudes: $\sin^2 2\theta = 6.9 \times 10^{-4}$ (top left), $\sin^2 2\theta = 3 \times 10^{-4}$ (top right), and $\sin^2 2\theta = 2 \times 10^{-3}$ (bottom).
\label{fig:app_allowedregions}}
\end{figure}

\begin{figure}[h]
\begin{center}
\includegraphics[width=0.4\columnwidth]{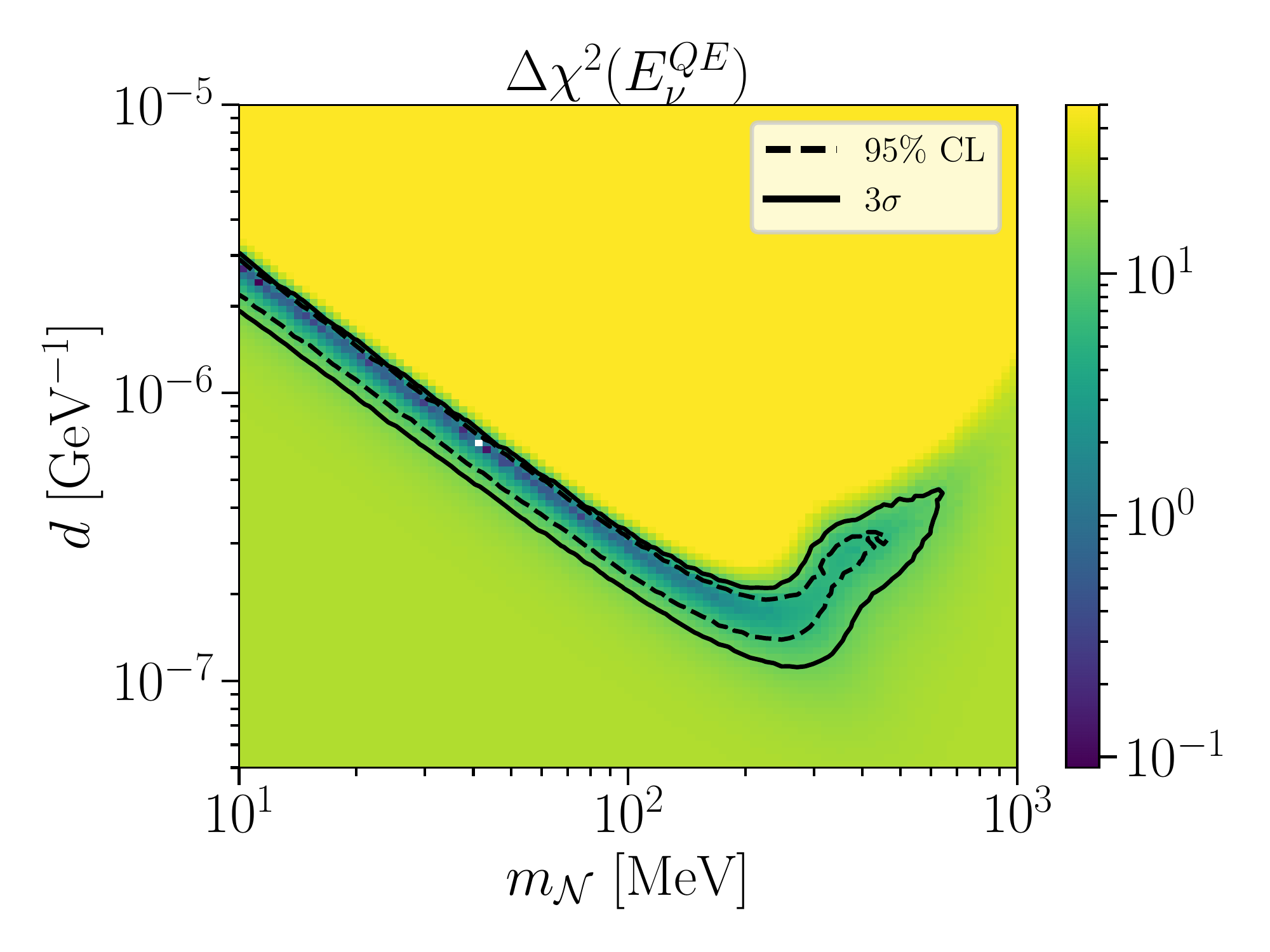}
\includegraphics[width=0.4\columnwidth]{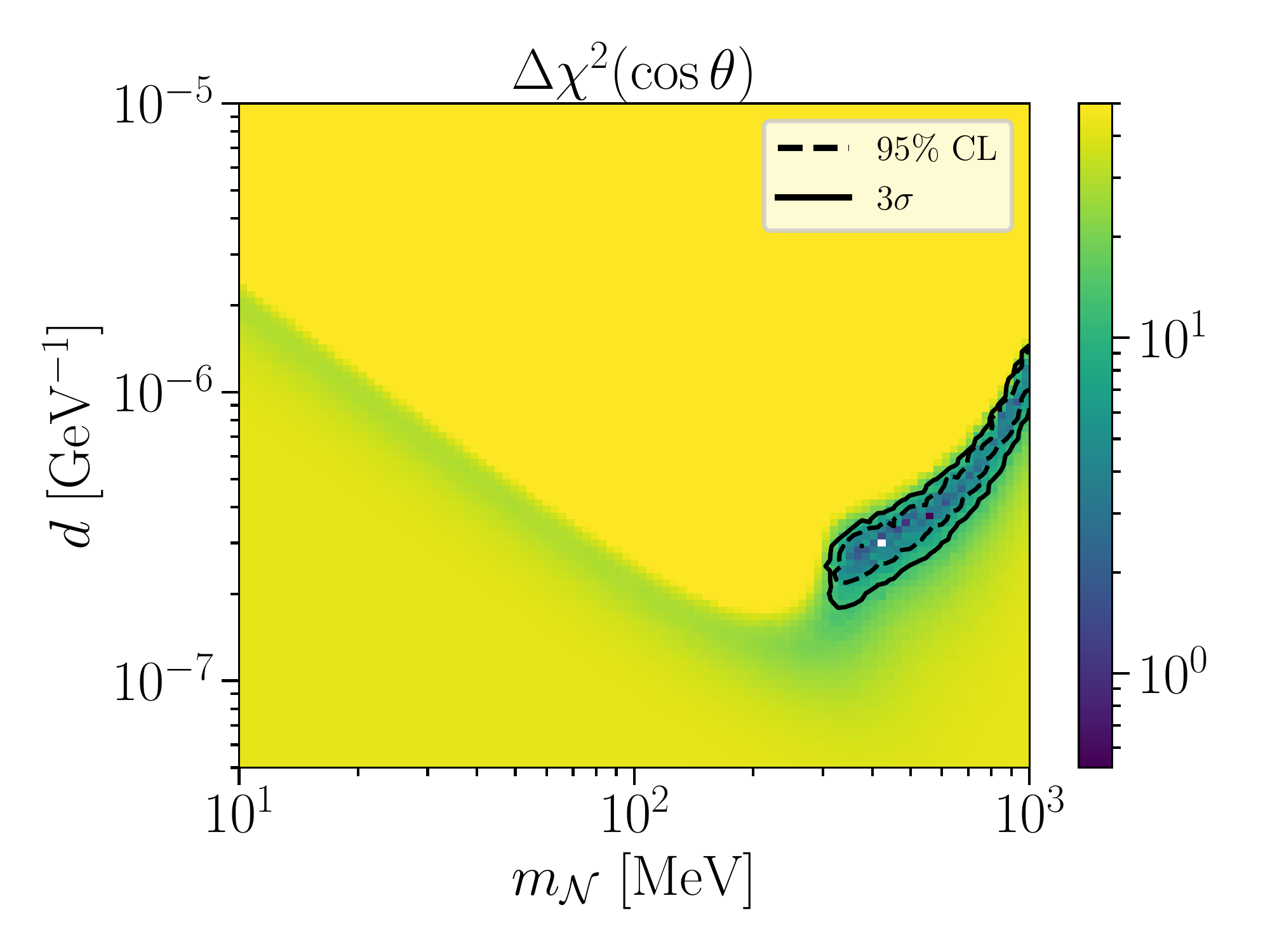}
\includegraphics[width=0.4\columnwidth]{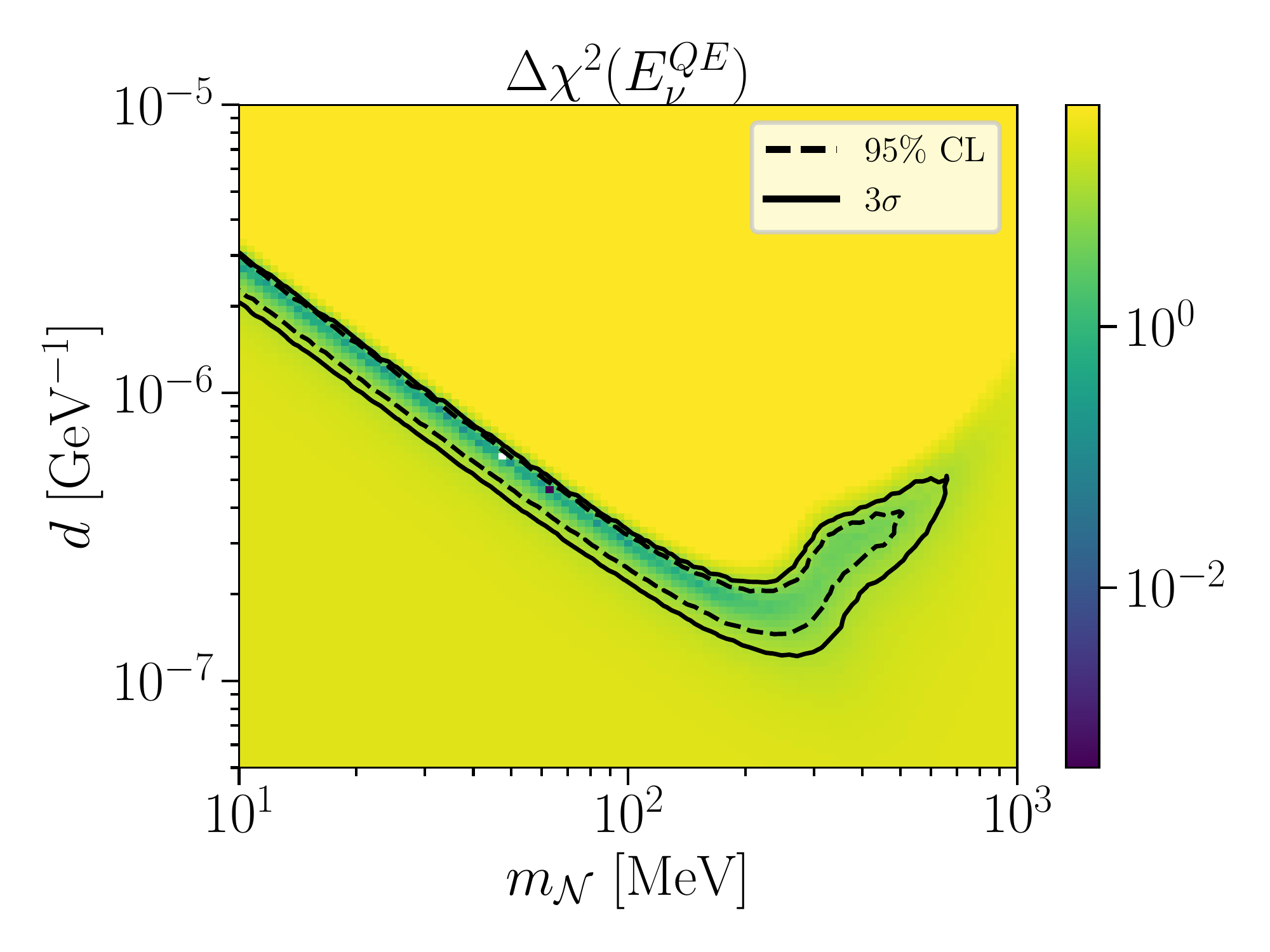}
\includegraphics[width=0.4\columnwidth]{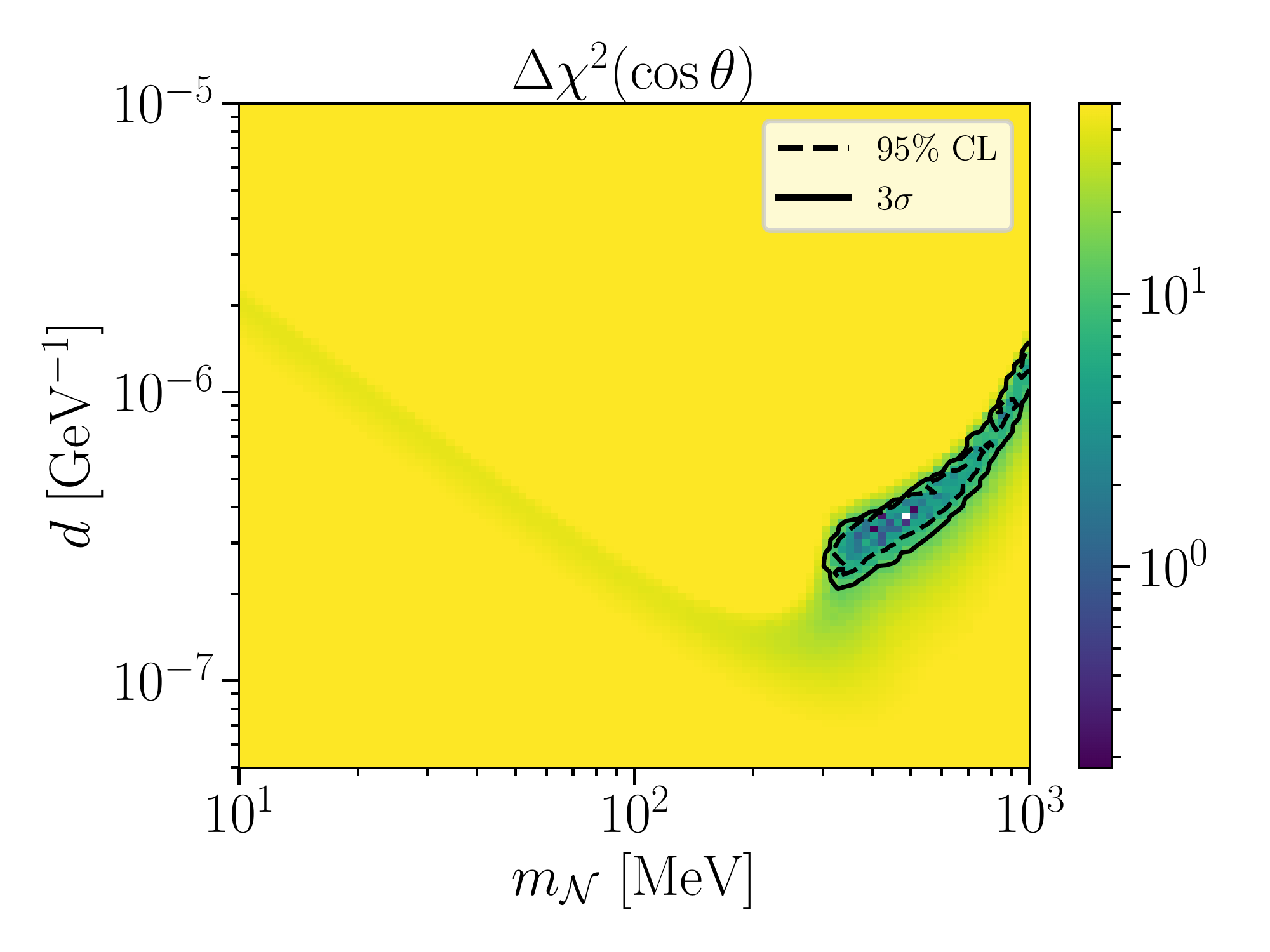}
\includegraphics[width=0.4\columnwidth]{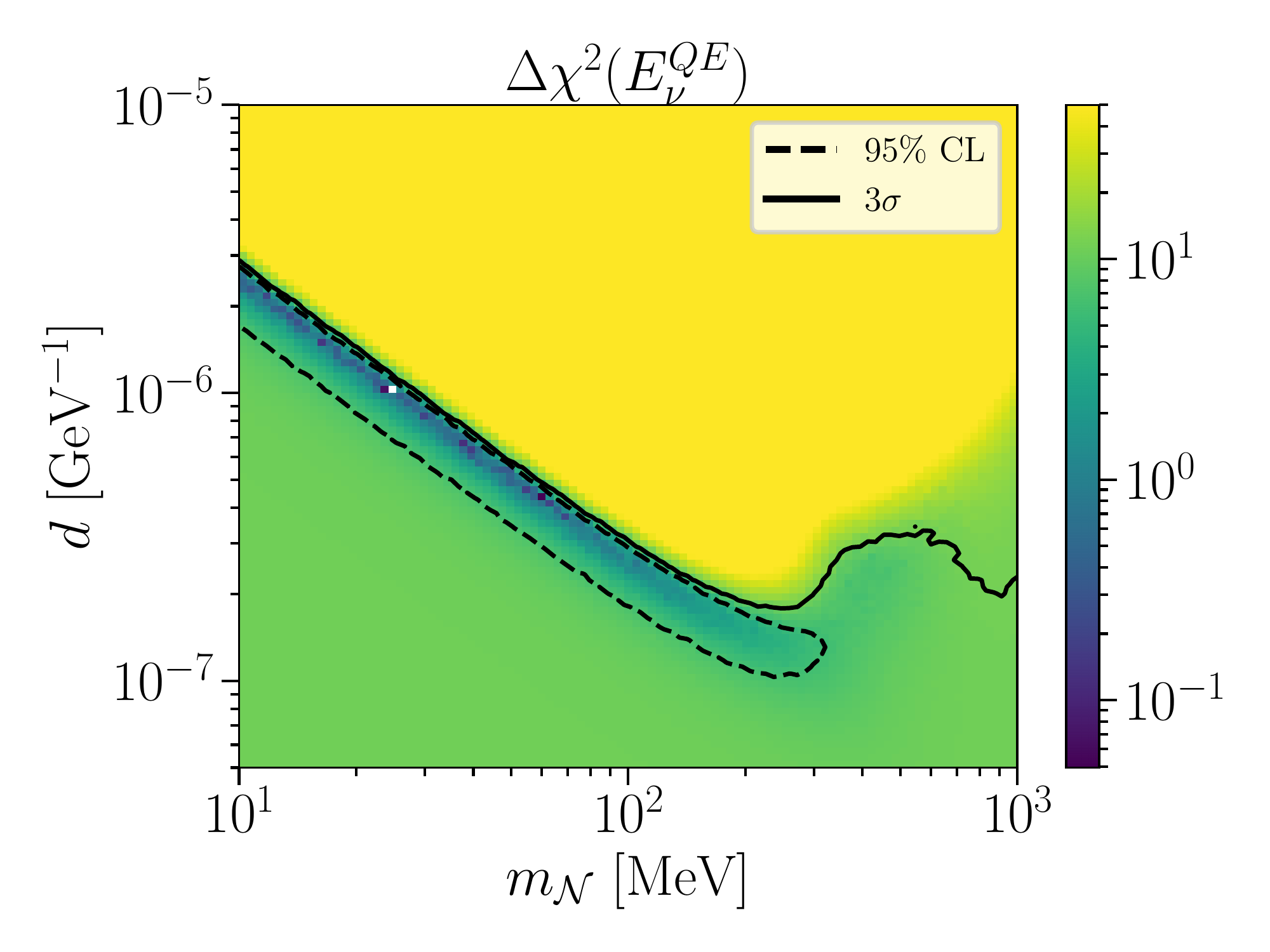}
\includegraphics[width=0.4\columnwidth]{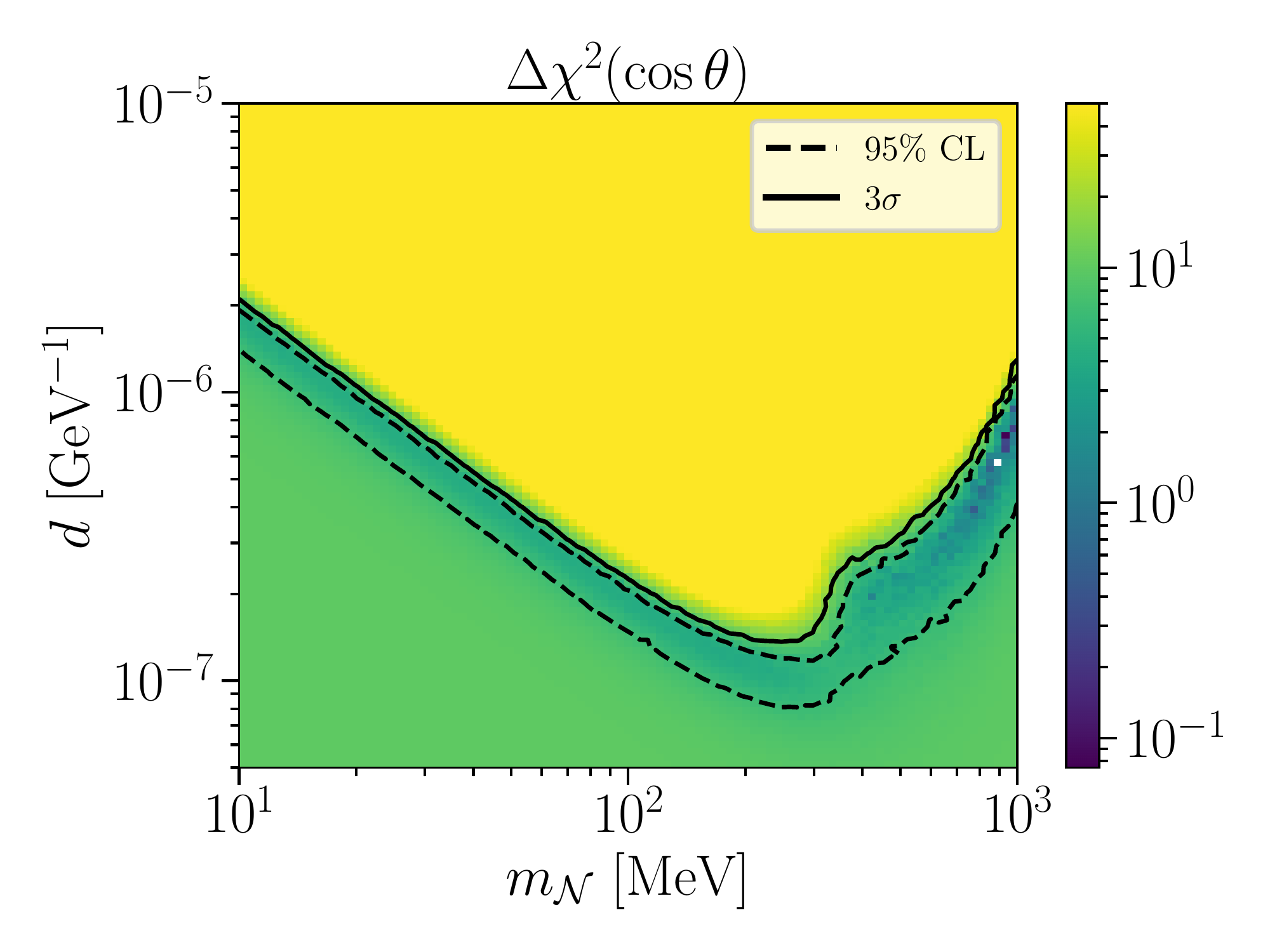}
\end{center}
\caption{Maps of $\Delta \chi^2 = \chi^2(d,m_\mathcal{N}) - \min_{\{d,\mathcal{N}\}} \chi^2(d,m_\mathcal{N})$ in HNL dipole model parameter space for the energy (left) and angular (right) fits to the MiniBooNE excess, considering an oscillation contribution with $\Delta m^2 = 1.3$ eV$^2$ and three different amplitudes: $\sin^2 2\theta = 6.9 \times 10^{-4}$ (top), $\sin^2 2\theta = 3 \times 10^{-4}$ (middle), and $\sin^2 2\theta = 2 \times 10^{-3}$ (bottom).
\label{fig:app_delchi}}
\end{figure}

\begin{figure}[h]
\begin{center}
\includegraphics[width=0.4\columnwidth]{figures/FinalPlotsNick/Eplot_02.pdf}
\includegraphics[width=0.4\columnwidth]{figures/FinalPlotsNick/UZplot_02.pdf}
\includegraphics[width=0.4\columnwidth]{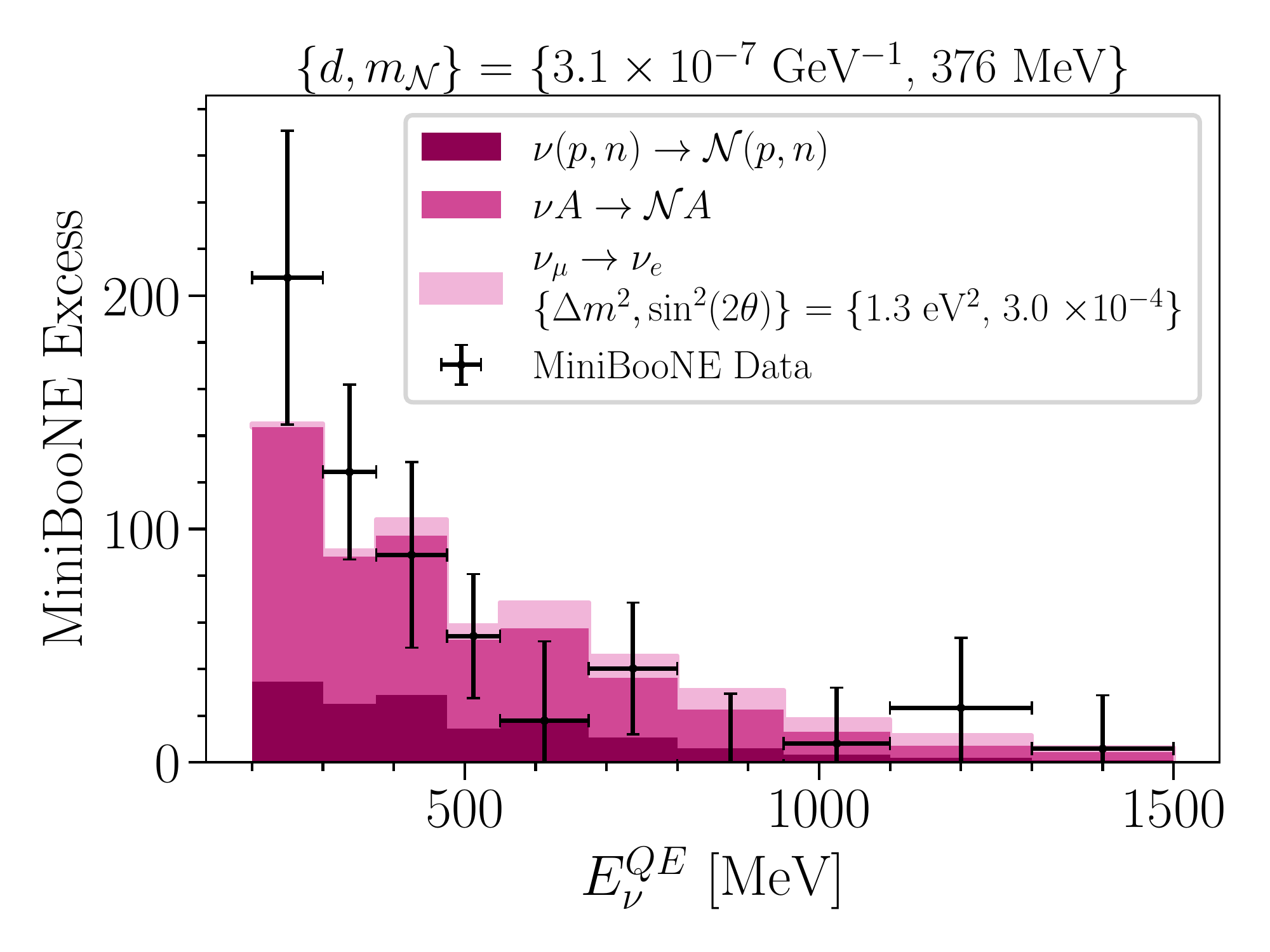}
\includegraphics[width=0.4\columnwidth]{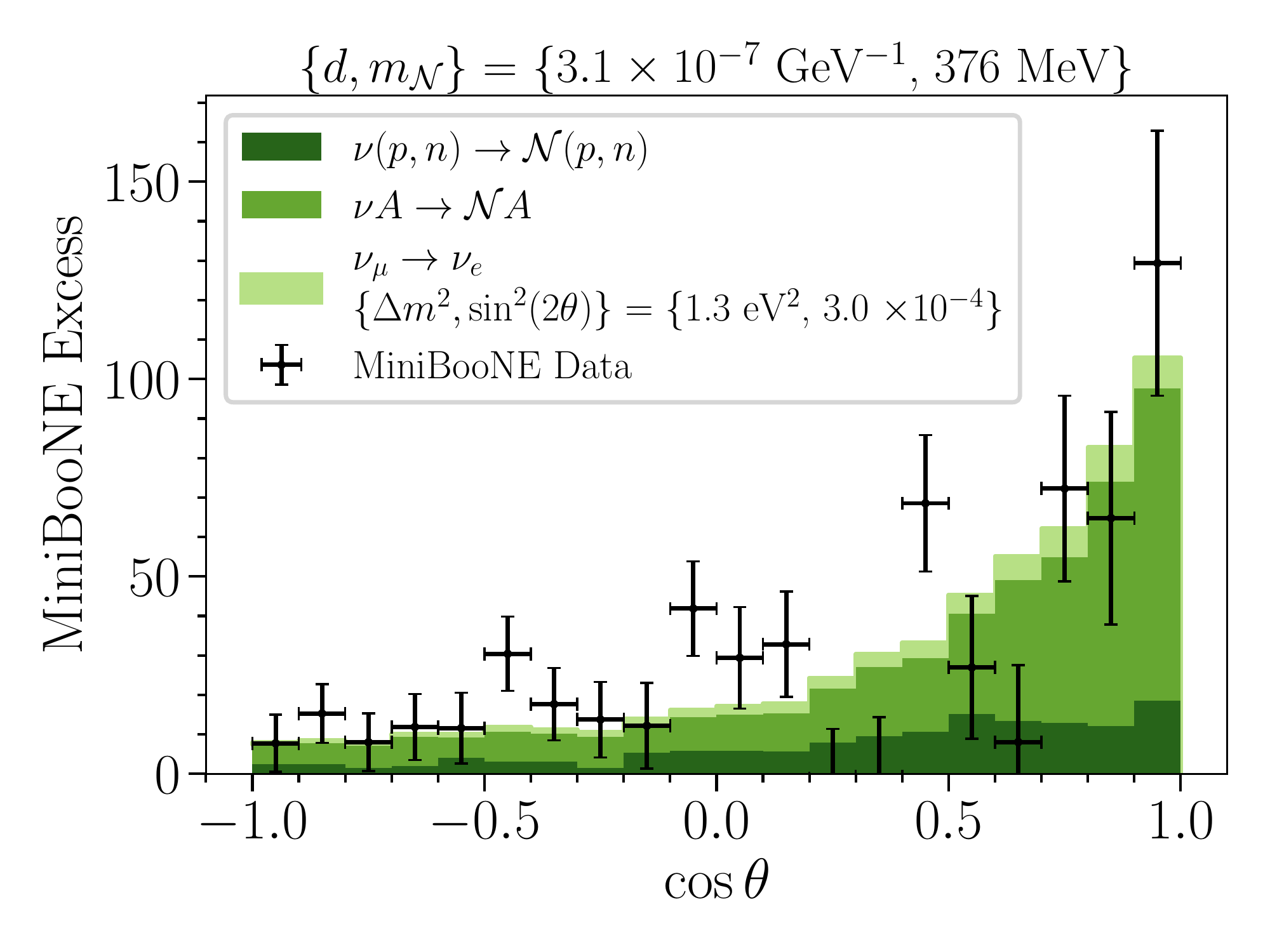}
\includegraphics[width=0.4\columnwidth]{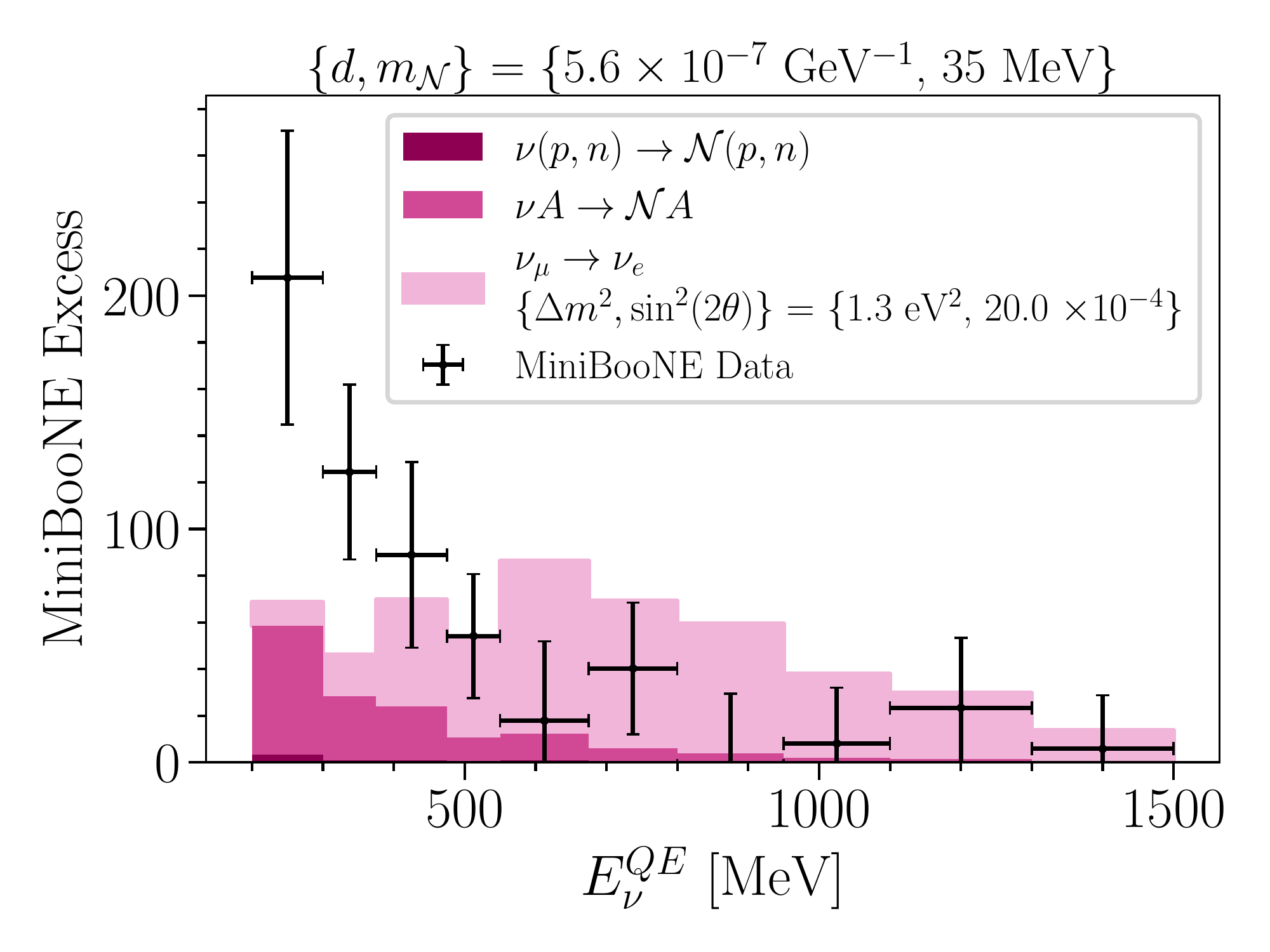}
\includegraphics[width=0.4\columnwidth]{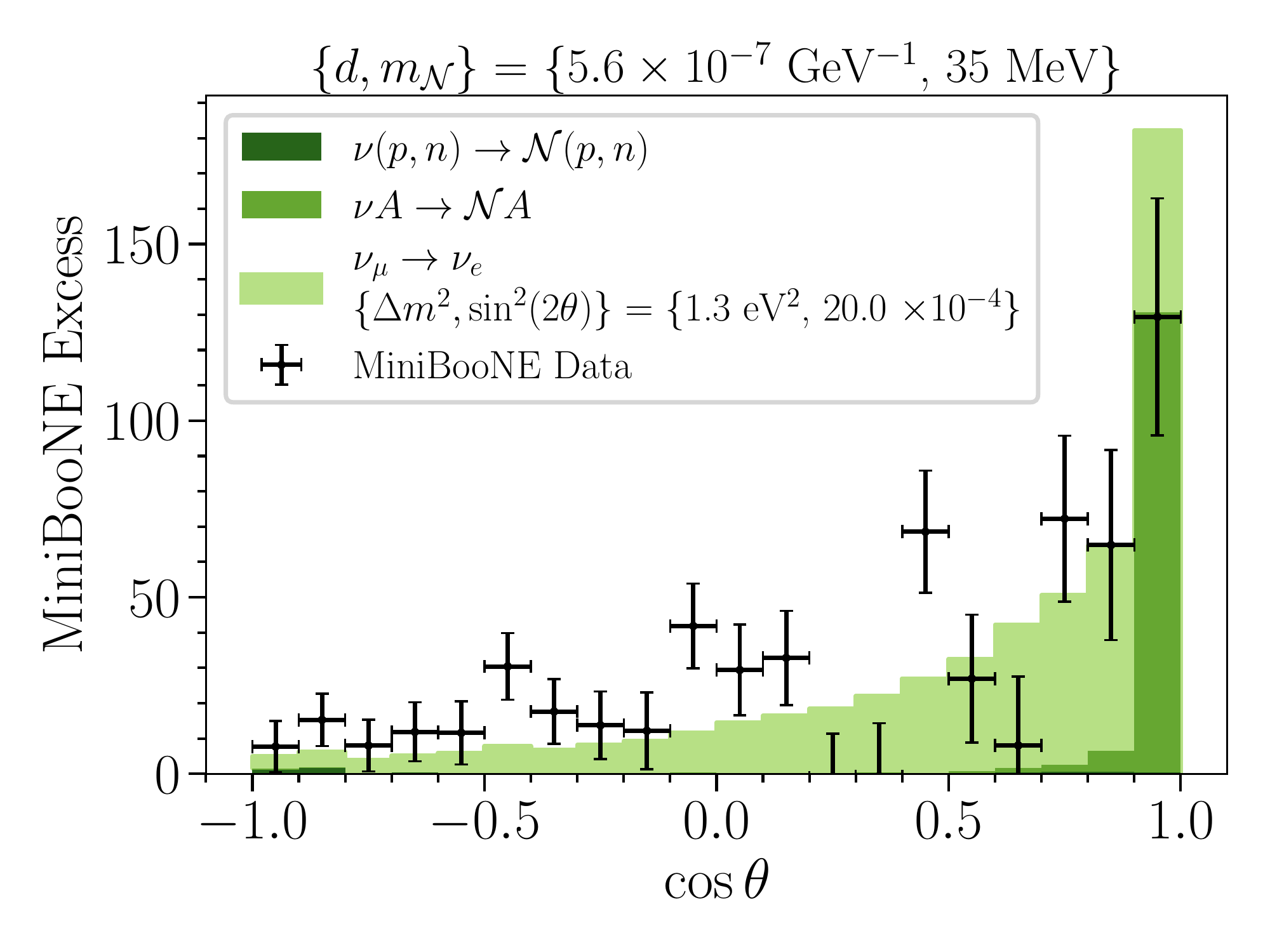}
\end{center}
\caption{Example HNL dipole model predictions for the energy (left) and angular (right) distributions compared with the MiniBooNE excess, considering an oscillation contribution with $\Delta m^2 = 1.3$ eV$^2$ and three different amplitudes: $\sin^2 2\theta = 6.9 \times 10^{-4}$ (top), $\sin^2 2\theta = 3 \times 10^{-4}$ (middle), and $\sin^2 2\theta = 2 \times 10^{-3}$ (bottom). The specific $d$ and $m_\mathcal{N}$ values shown here come from Table~\ref{tab:fitchisq}.
\label{fig:app_dists}}
\end{figure}

Figure~\ref{fig:app_allowedregions} shows the energy/angular fit $95\%$ CL allowed regions in $\{d,m_\mathcal{N}\}$ parameter space for the three $\sin^2 2\theta$ values considered in Table~\ref{tab:fitchisq}.
One can see that the overlap region between the two fits becomes less strict for larger oscillation contributions.
This is partially because larger oscillation contributions give a poorer fit to the MiniBooNE excess, as shown by the $\Delta \chi^2$ plots in Figure~\ref{fig:app_delchi}.
Here one can see that for the largest $\sin^2 2\theta$, there is no longer a closed contour in either the $E_\nu^{QE}$ or $\cos \theta$ fit at $3\sigma$ CL.
Figure~\ref{fig:app_dists} shows each of the $E_\nu^{QE}$ and $\cos \theta$ predictions from Table~\ref{tab:fitchisq} compared with the MiniBooNE excess.
These plots further indicate a preference for a smaller $eV$-scale oscillation contribution (considering global best-fit oscillation parameters) in order to fit MiniBooNE. 

\begin{figure}[h]
\begin{center}
\includegraphics[width=0.4\columnwidth]{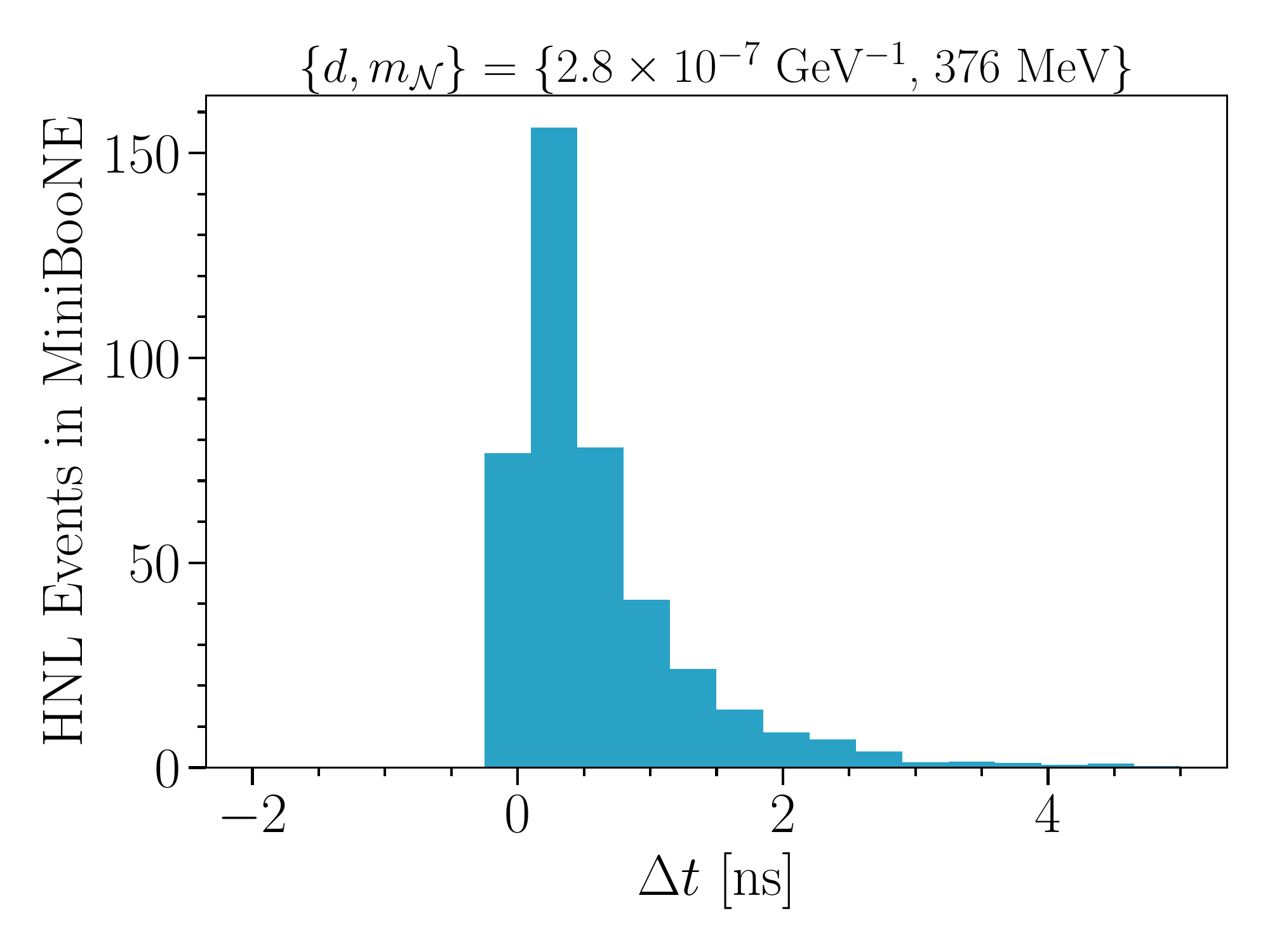}
\includegraphics[width=0.4\columnwidth]{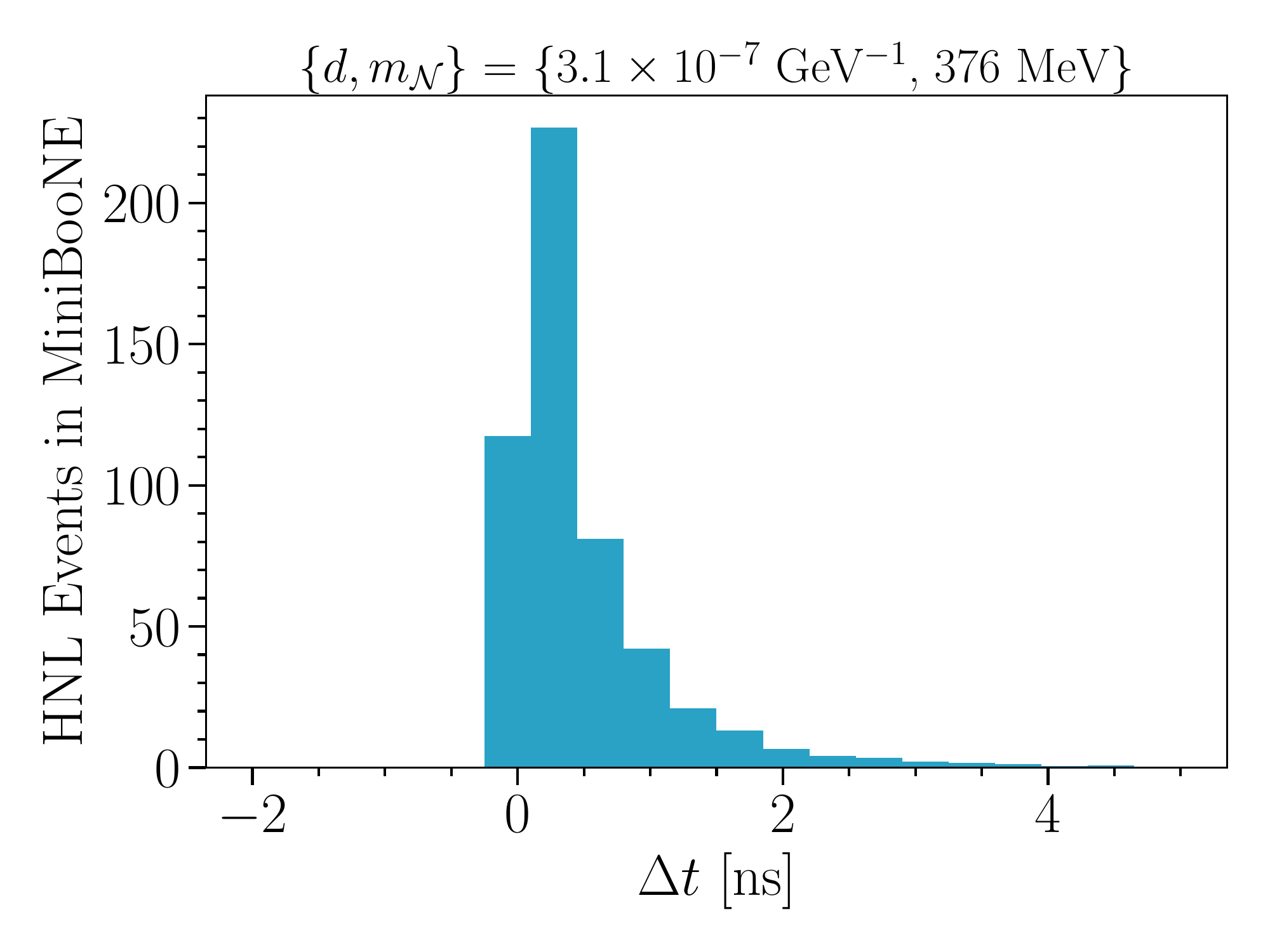}
\includegraphics[width=0.4\columnwidth]{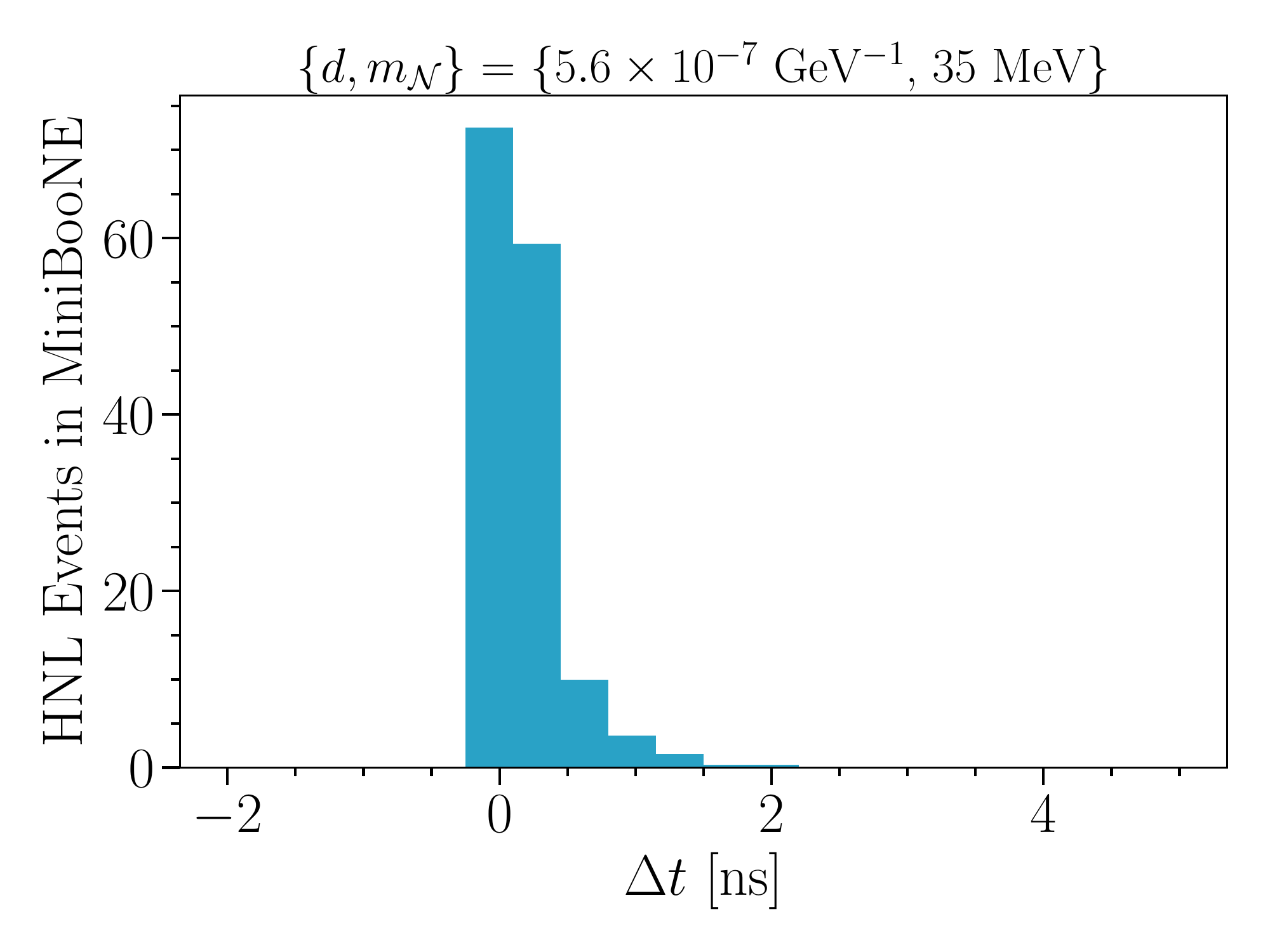}
\end{center}
\caption{Timing delay distributions for HNL events in the MiniBooNE detector, as defined in the text. The specific $d$ and $m_\mathcal{N}$ values shown here come from Table~\ref{tab:fitchisq}. Three different $eV$-scale oscillation amplitudes are considered: $\sin^2 2\theta = 6.9 \times 10^{-4}$ (top left), $\sin^2 2\theta = 3 \times 10^{-4}$ (top right), and $\sin^2 2\theta = 2 \times 10^{-3}$ (bottom).  
\label{fig:app_tplot}}
\end{figure}

Finally, we consider the timing delay distribution for HNL decays in the MiniBooNE detector.
The timing delay is defined as the time between HNL production and decay minus the same time it would take for a speed of light neutrino to travel the same distance.
Figure~\ref{fig:app_tplot} shows this delay for the three different oscillation amplitude cases, indicating timing delays small enough to be consistent with the MiniBooNE excess timing distribution~\cite{MiniBooNE:2020pnu}.

\subsection*{Meson Decay Simulation}

The $\pi^0$s have been simulated using a Sanford-Wang (SW) parametrization with $\pi^0$ distribution calculated as the average between the $\pi^+$ and $\pi^-$ distributions with coefficients taken from~\cite{PhysRevD.79.072002}.
A rejection sampling from previous studies~\cite{deNiverville:2017} has been used to simulate momenta $p_{\pi^0}\in [0, 8] \:$ GeV, emission angles with respect to the direction of the beam $\theta_{\pi^0} \in [0, \frac{\pi}{2}]$, and angle $\phi_{\pi^0} \in [0,2\pi]$ of $\pi^0$ produced at the source in lab frame.
Four-momentum vectors have been created with the aforementioned triplets $p_{\pi^0}$, $\theta_{\pi^0}$, and $\phi_{\pi^0}$ according to $p^\mu_{\pi^0}=(E_{\pi^0},p_{\pi^0}[\cos(\phi_{\pi^0})\sin(\theta_{\pi^0}),\sin(\phi_{\pi^0})\sin(\theta_{\pi^0}),\cos(\theta_{\pi
^0})])$.
A decay probability $P_{decay}$ has been associated to each $\pi^0$ according to Eq.~\ref{eq:decaylength}.
In this study, $\mathcal{N}$ have been created via the three body Dalitz-like decay $\pi^0 \rightarrow \mathcal{N}+\nu+\gamma$. Assuming $\nu$ to be massless, the energy $E_{\mathcal{N}}$ in the $\pi^0$ rest frame has been constrained to be $E_{\mathcal{N}} \in (m_\mathcal{N},\frac{{m_{\pi^0}}^2+{m_\mathcal{N}}^2}{2m_{\pi^0}})$ and the differential branching ratio in the $\pi^0$ rest
frame for each $\mathcal{N}$ has been evaluated using Eq.\ A.10 in~\cite{Magill:2018jla}.
The simulated $\mathcal{N}$ have been subsequently boosted to the lab frame of the $\pi^0$.
Considering the channel $\mathcal{N} \rightarrow \gamma + \nu$ with a massless $\nu$, the $\gamma$ has been simulated assuming their energies $E_\gamma = \frac{m_{\mathcal{N}}}{2}$, emission angle $cos(\theta_\gamma) \in [-1,1]$, angle $\phi_\gamma \in [0,2\pi]$, and associated four-momentum vectors to be $p^\mu_{\gamma}=(E_{\gamma},p_{\gamma}[\cos(\phi_{\gamma})\sin(\theta_{\gamma}),\sin(\phi_{\gamma})\sin(\theta_{\gamma}),\cos(\theta_\gamma)])$.
The resulting $\gamma$s have then been boosted to the lab frame of the $\mathcal{N}$ and multiplied by a normalization constant $N_{const}=\frac{N_{\pi^0}}{N_{sim}}$. $N_{\pi^0}=1.17*10^{21}$ is the number of $\pi^0$ produced in neutrino mode over the lifetime of MiniBooNE for $18.75*10^{20}$ POT~\cite{MiniBooNE:2020pnu} and a $\pi^0$ multiplicity per POT 0.9098~\cite{Magill:2018jla}. $N_{sim}$ is the effective number of $\pi^0$ decays actually simulated.

\section{Further Discussion on the Fits}

\begin{table}[b]
    \centering
    \begin{tabular}{|c|c|c|c|c|}
        \hline
        Experiment & Ref. & Type & In $3+1+\mathcal{N}$ osc fit? & In $3+1+\mathcal{N}$ decay fit? \\  
         \hline
         Bugey & \cite{Bugey} & $\bar \nu_e \rightarrow \bar \nu_e$ & Yes & No \\
         NEOS  & \cite{NEOS} & $\bar \nu_e \rightarrow \bar \nu_e$  & Yes & No\\
         DANSS  & \cite{DANSS} & $\bar \nu_e \rightarrow \bar \nu_e$ & Yes & No \\
         PROSPECT  & \cite{PhysRevD.103.032001} & $\bar \nu_e \rightarrow \bar \nu_e$  & Yes & No \\
         STEREO  & \cite{PhysRevD.102.052002} & $\bar \nu_e \rightarrow \bar \nu_e$  & Yes & No\\ \hline
         KARMEN/LSND Cross Section & \cite{ConradShaevitz} & $\nu_e \rightarrow \nu_e$  & Yes & No\\
         Gallium & \cite{Gallex, SAGE} & $\nu_e \rightarrow \nu_e$  & Yes & No \\ \hline
         SciBooNE/MiniBooNE & \cite{SBMBnubar} & $\bar \nu_\mu \rightarrow \bar \nu_\mu$  & Yes & No \\        
         CCFR & \cite{CCFR84} & $\bar \nu_\mu \rightarrow \bar \nu_\mu$ & Yes & No\\        
         MINOS & \cite{MINOSCC2012,MINOSCC2011} & $\bar \nu_\mu \rightarrow \bar \nu_\mu$ & Yes & No \\ \hline        
         SciBooNE/MiniBooNE & \cite{SBMBnu} & $\nu_\mu \rightarrow \nu_\mu$  & Yes & No\\        
         CCFR & \cite{CCFR84} & $\nu_\mu \rightarrow \nu_\mu$ & Yes & No \\         
         CDHS & \cite{CDHS} & $\nu_\mu \rightarrow \nu_\mu$  & Yes & No \\  
         MINOS & \cite{MINOS2016} & $\nu_\mu \rightarrow \nu_\mu$  & Yes & No \\ \hline        
         LSND & \cite{Aguilar:2001ty} & $\bar \nu_\mu \rightarrow \bar \nu_e$  & Yes & No\\        
         KARMEN & \cite{KARMEN} & $\bar \nu_\mu \rightarrow \bar \nu_e$  & Yes & No\\        
         MiniBooNE (BNB) & \cite{MBnubar} & $\bar \nu_\mu \rightarrow \bar \nu_e$ & No & No \\ \hline 
         NOMAD & \cite{NOMAD1} & $\nu_\mu \rightarrow  \nu_e$ & Yes & No\\        
         MiniBooNE (NuMI) & \cite{MBNumi} & $\nu_\mu \rightarrow \nu_e$ & No & No \\        
         MiniBooNE (BNB) & \cite{PhysRevD.103.052002} & $\nu_\mu \rightarrow \nu_e$ & No & Yes\\ \hline 
    \end{tabular}
    \caption{Table of experimental results used in the fits in this paper.}
    \label{tab:expts}
\end{table}

This Appendix provides further information on the fit results presented in this paper.   

Table~\ref{tab:expts} provides a table of the experimental results used in the fits, with references. 
Notation in Column 3 refers to electron flavor disappearance, muon flavor disappearance, and muon-to-electon flavor appearance for neutrinos and antineutrinos.
The lastt two columns indicate the data sets used in the fits to the new $3+1+\mathcal{N}$ model.
This model has two contributions: the oscillation contribution, established through a 3+1 oscillation fit to experiments in Column 3 and the $\mathcal{N}$ decay fit which is established through a fit to the MiniBooNE 2020 data set only, as indicated in Column 4.
This omits MiniBooNE (BNB) antineutrino data and MiniBooNE (NuMI) data from the $\mathcal{N}$ decay fit, which have anomalous signals consistent with MiniBooNE (BNB) neutrino data, but at low statistics and with high backgrounds.
Including these requires additional development of Monte Carlo beamline simulations, and the modest results will not change the conclusion of this paper, thus we conclude this is beyond the scope of this article.

Below, we provide the results of the oscillations-only global fits using the data indicated in Table~\ref{tab:expts}. 
A thorough description of the fitting process is provided in Ref.~\cite{Diaz:2019fwt}.
In addition to the experiments listed in Ref.~\cite{Diaz:2019fwt}, we have updated the MiniBooNE (BNB) neutrino appearance data set~\cite{PhysRevD.103.052002}, updated the PROSPECT antineutrino disappearance data set~\cite{PhysRevD.103.032001}, and introduced the STEREO antineutrino disappearance data set. 

In reporting the tension, we make use of the probability associated with the Parameter Goodness of Fit Test, or PGF Test, which is a standard in our field.
In this test~\cite{Maltoni:2003cu}, the global (glob) set data is divided into disappearance (dis) and appearance (app) sets.
One defines the $\chi^2$ and degrees of freedom as:
\begin{eqnarray}
\chi^2_{PGF} & =& \chi^2_{glob}-(\chi^2_{app}+\chi^2_{dis}),  \label{chi2pg}\\
N_{PGF} &= &(N_{app}+N_{dis}) - N_{glob}~.   \label{npg}
\end{eqnarray}
We quote the associated probability as the tension.   Table~\ref{table:fitquality} lists the inputs to the calculation that appears in this paper.

\begin{figure*}[tb]
\begin{center}
\includegraphics[width=0.3\columnwidth]{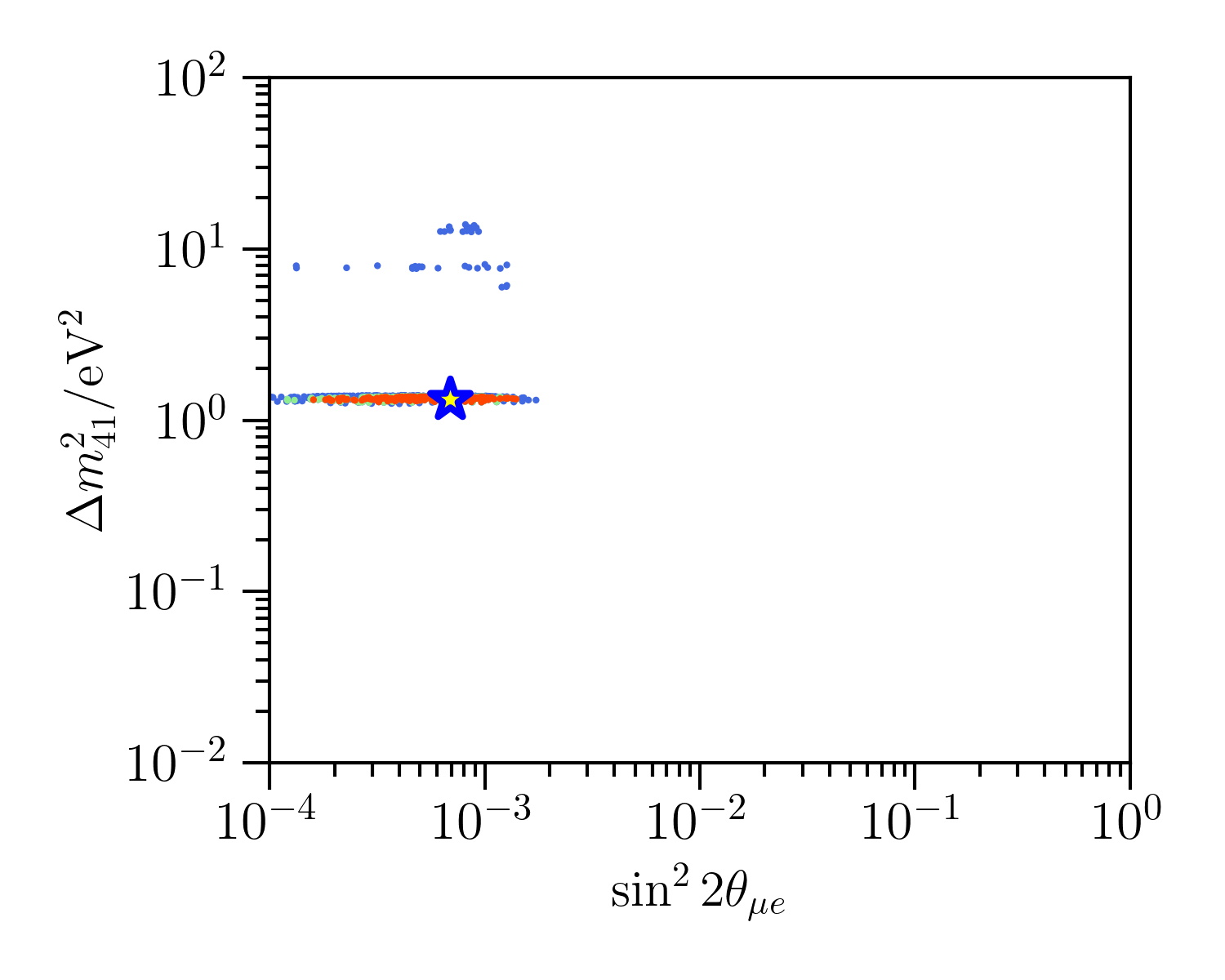}
\includegraphics[width=0.3\columnwidth]{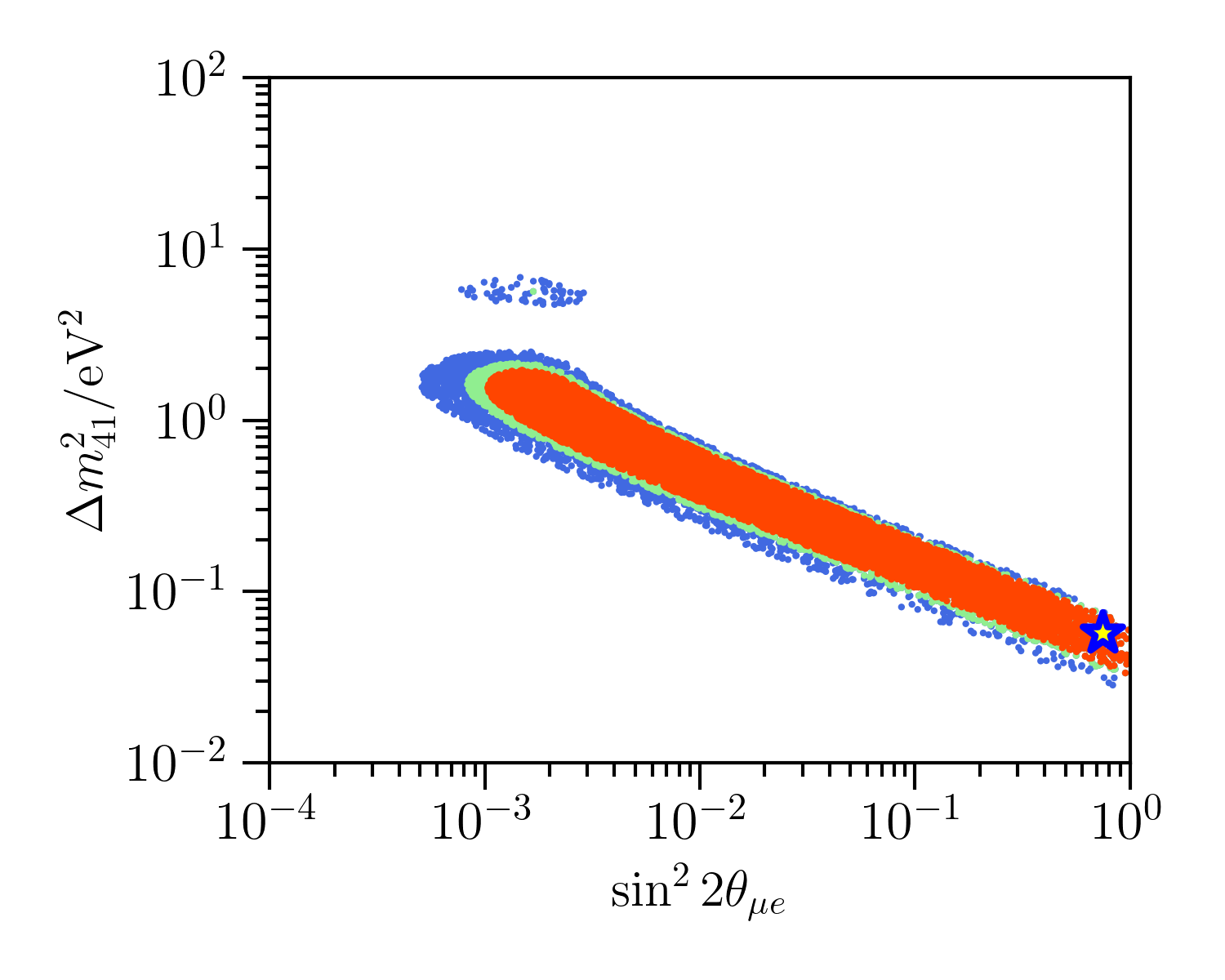}
\includegraphics[width=0.3\columnwidth]{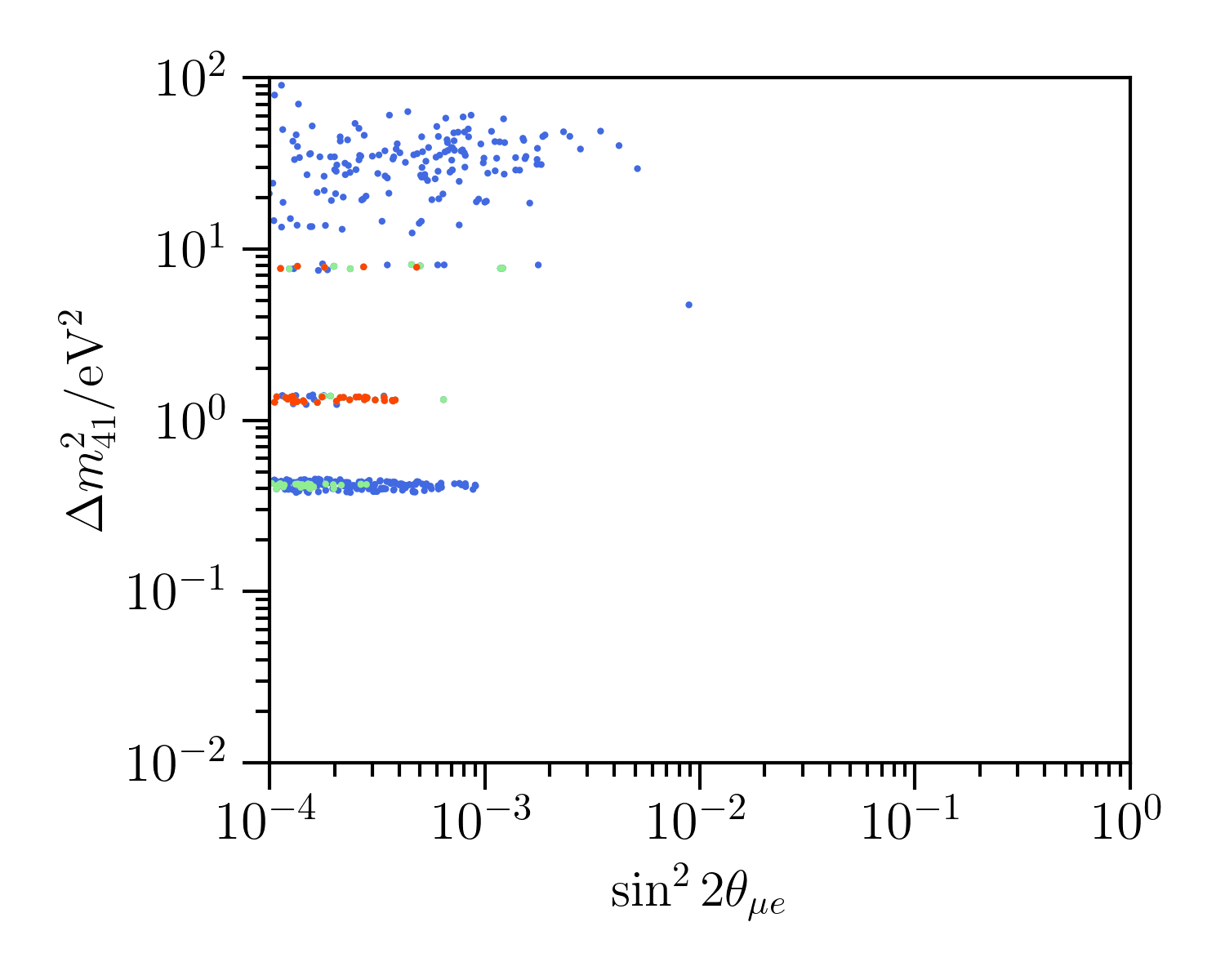}
\end{center}
\caption{These plots show the best fit regions for the 3+1+only oscillation data sets used in this work. The left plot shows the best fit regions of the global fits, with 90\%, 95\%, and 99\% regions shown in red, green, and blue, respectively. The middle plot shows the best fit region when only fitting to the appearance subset of data, while the right plot shows the best fit region when only fitting to the disappearance subset. The tension within this data set can be qualitatively seen by the fact that the best fit regions do not overlap between the appearance-only and disappearance-only data sets. 
\label{fig:3plus1withoutMBNUMI}}
\end{figure*}

For the 3+1-only fit used in this article, we run the oscillation fit on the experiments that would not be sensitive to the $\mathcal{N}$ decay.
As stated earlier, this procedure finds a best fit point of $\Delta m^2 = 1.32$ eV$^2$ and $\sin^2 2\theta_{e\mu} = 6.9 \times 10^{-4}$, with an allowed range $\sin^2 2 \theta_{\mu e} \in [3\times 10^{-4},2 \times 10^{-3}]$ at the 90\% CL.
The best-fit regions are shown on the leftmost plot in Fig~\ref{fig:3plus1withoutMBNUMI}.
The tension within this fit is demonstrated by separately fitting the appearance and disappearance data sets, which are shown in the middle and right plot in~\ref{fig:3plus1withoutMBNUMI}, respectively.
This tension is found to have a p-value of $7\times 10^{-3}$, with the relevant parameters summarized in Table~\ref{table:fitquality}. 

\begin{figure*}[tb]
\begin{center}
\includegraphics[width=.3\columnwidth]{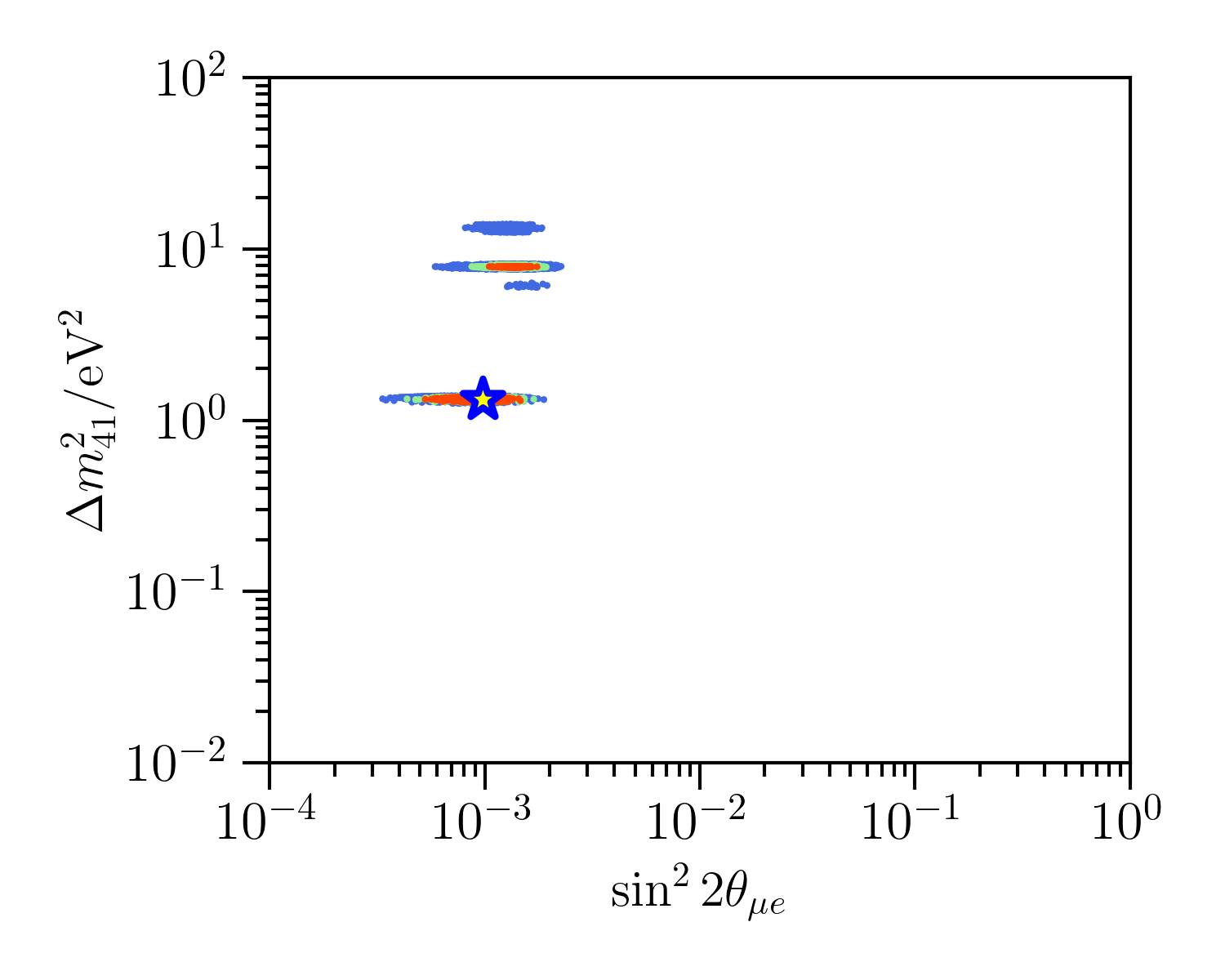}
\includegraphics[width=.3\columnwidth]{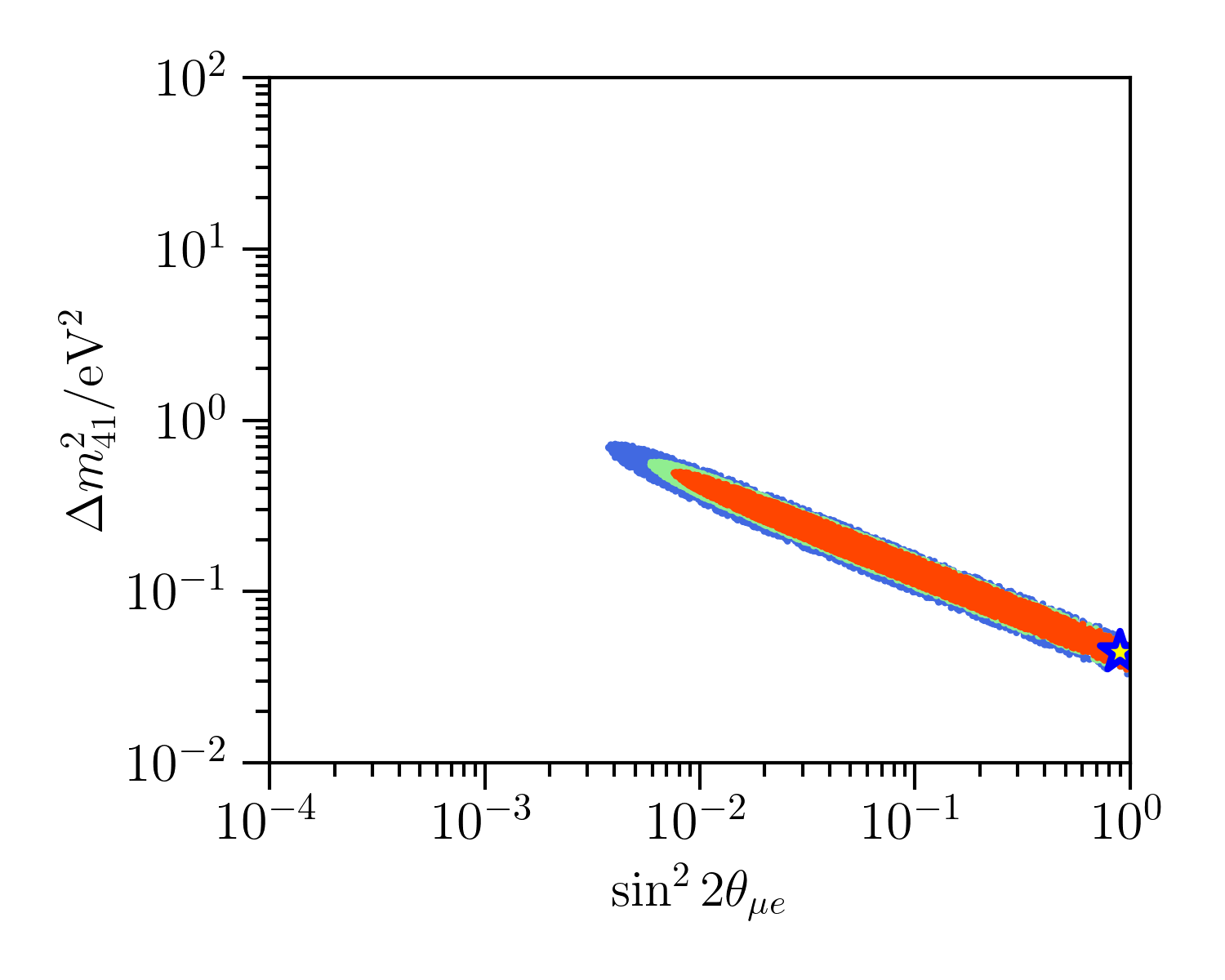}
\includegraphics[width=.3\columnwidth]{figures/disappearance.png}
\end{center}
\caption{These plots show the best fit regions for the 3+1+complete oscillation data sets. This serves as an update to the fits provided in \cite{Diaz:2019fwt}, and are not used in the preceding analysis. The left plot shows the best fit regions of this complete global fit, with 90\%, 95\%, and 99\% regions shown in red, green, and blue, respectively. The middle plot shows the best fit region for the appearance subset of data, while the right plot shows the best fit region for the disappearance subset. As seen before, the global sterile neutrino oscillation model suffers from substantial tension.
\label{fig:3plus1}}
\end{figure*}

For completeness, we also provide a global fit including \textit{all} experiments listed in Table~\ref{tab:expts}.
These results, which we'll label ``3+1-complete," are not used in the preceding analysis.
The best fit regions for this $3+1$ global fit is shown in the left plot of  Fig.~\ref{fig:3plus1}, with the best fit point at $\sin^2(2\theta_{\mu e}) = 1 \times 10^{-3}$ and $\Delta m_{41}^2 = 1.32$ eV$^2$.
As has been seen before~\cite{Collin:2016aqd,Dentler:2018sju,Diaz:2019fwt,Boser:2019rta}, this model suffers from a tension between the appearance and disappearance data sets.
The best-fit regions of the appearance and disappearance data sets are shown in the middle and right plot of Fig.~\ref{fig:3plus1}, respectively.
The tension found in the global data set is found to have a p-value of $8\times 10^{-7}$, with the relevant parameters summarized in Table~\ref{table:fitquality}.

\begin{figure*}[tb]
\begin{center}
\includegraphics[width=6.in]{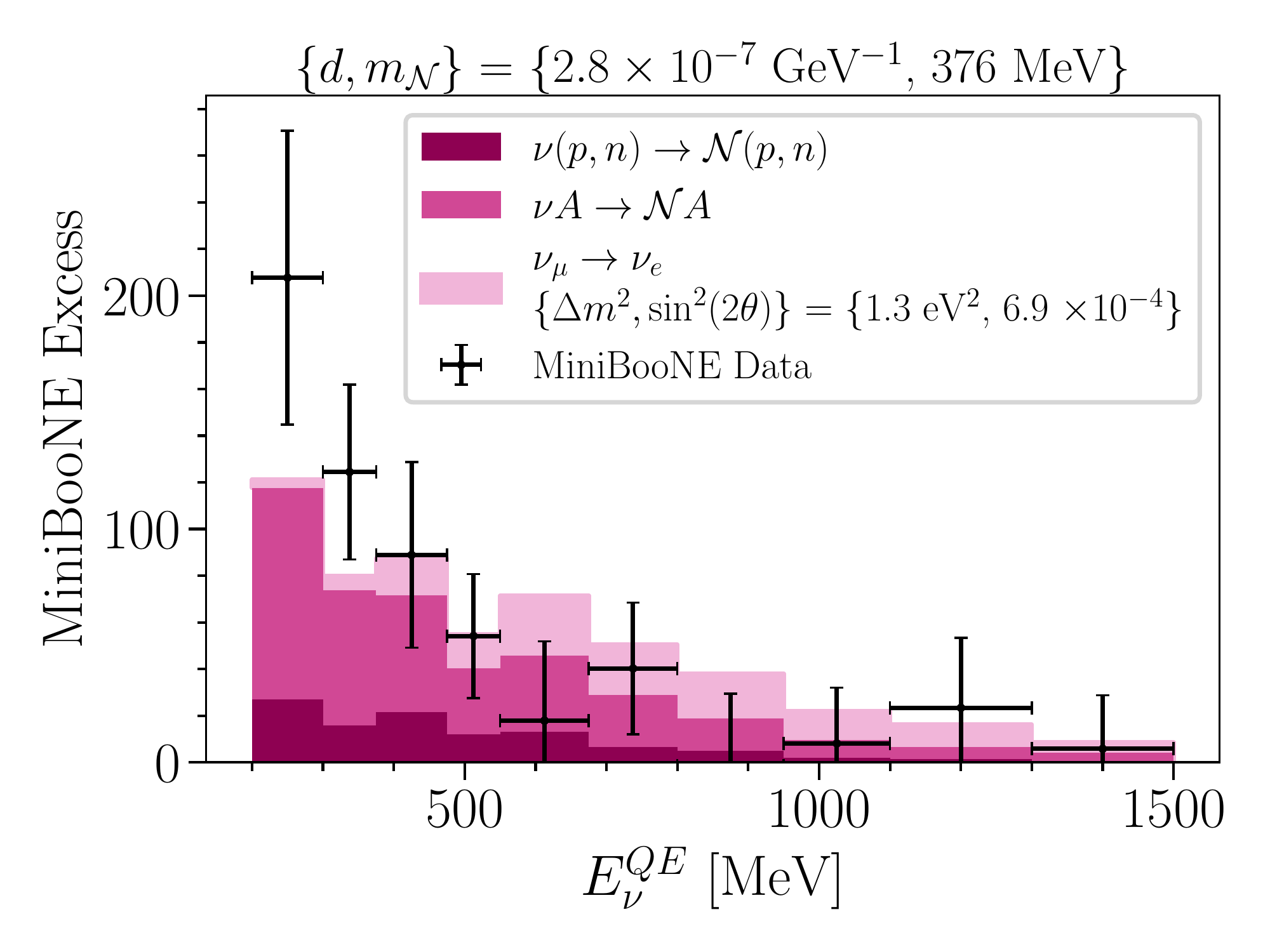}
\end{center}
\caption{Enlarged version of Fig.~\ref{fig:energyangle}, left, in the paper.  The prediction for the oscillation contribution to the $3+1+\mathcal{N}$ component is shown in light pink.   The prediction for the $\mathcal{N}$ decay component, from both coherent and incoherent upscattering, is shown in darker pink.  See text in this appendix for explanation of the data.
\label{fig:energy}}
\end{figure*}

Our analysis fits to the MiniBooNE (BNB) neutrino energy distribution, which has been presented with statistical and systematic error.
Specifically, the points and errors are taken from Fig.~19 of Ref.~\cite{PhysRevD.103.052002}.
In this case, the data are the excess events when the MiniBooNE measurement is compared to the constrained backgrounds.
It is important to note that for most neutrino energy bins, the systematic error dominates, and so is necessary to include in the uncertainty when performing the fit.
The result of the $3+1+\mathcal{N}$ energy fit has been shown in the paper in Fig.~\ref{fig:energyangle}, left, and here we reproduce and enlarge the same figure in Fig.~\ref{fig:energy}.
Light pink indicates the oscillation component, with $\Delta m^2=1.3$ eV$^2$ and $\sin^2 2\theta=6.9 \times 10^{-4}$. The darker pink regions indicate the HNL decay component from both coherent and incoherent upscattering production, with $d=2.8 \times 10^{-7}$ GeV$^{-1}$ and $m_\mathcal{N} = 376$ MeV.

\begin{figure*}[tb]
\begin{center}
\includegraphics[width=6.in]{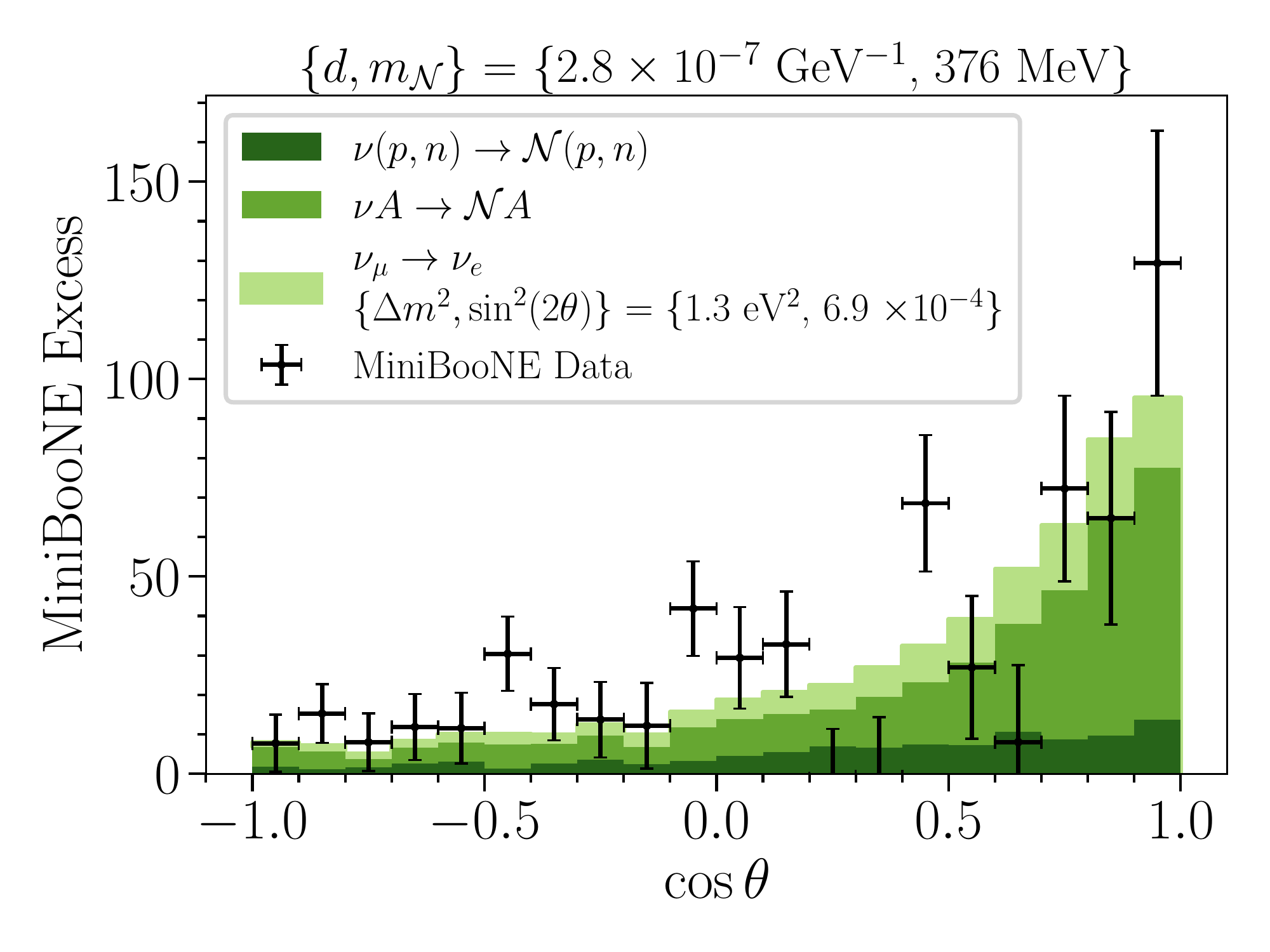}
\end{center}
\caption{Enlarged version of Fig.~\ref{fig:energyangle}, right, in the paper.  The prediction for the oscillation contribution to the $3+1+\mathcal{N}$ component is shown in light green.   The prediction for the $\mathcal{N}$ decay component, from both coherent and incoherent upscattering, is shown in darker green.  See text in this appendix for explanation of the data.
\label{fig:angle}}
\end{figure*}

MiniBooNE has not provided a data release for the angular distribution of the electromagnetic shower in the excess events. 
The data shown in Fig.~\ref{fig:energyangle}, right, is obtained by subtracting the unconstrained backgrounds from the MiniBooNE measurement shown in Fig.~8 of Ref.~\cite{PhysRevD.103.052002}. 
Only statistical errors are provided by MiniBooNE at present.
We note that one would expect the systematic uncertainty to be substantially larger, as was the case for the neutrino energy. Therefore, it would be interesting to repeat this analysis considering a robust treatment of the systematic errors in the angular distribution.
Fig.~\ref{fig:angle} reproduces and enlarges Fig.~\ref{fig:energyangle}, right, showing the oscillation component in light green and the HNL decay component from coherent/incoherent upscattering production in darker green.
One can see that the oscillation component does not address the forward-peak in the MiniBooNE data while the decay component primarily addresses that peak.

\begin{table*}
\begin{tabular}{|l | c | c | c |  }
\hline
Fit type:		&	3+1-only	&	3+1-complete \\ \hline
	$\chi^2_{app}$	&		48		&	79	\\
	$N_{app}$	&			2	&	2		\\
	$\chi^2_{dis}$	&		557		&	557		\\
	$N_{dis}$	&			3	&	3	\\
	$\chi^2_{glob}$	&			615	&	664		\\
	$N_{glob}$	&			3	&	3	\\
	$\chi^2_{PGF}$	&		10		&	28		\\
	$N_{PGF}$	&			2	&	2		\\
	$p$-value	&			7E-03	&	8E-07		\\
	$N\sigma$	&		2.5$\sigma$		&	4.8$\sigma$	 \\ \hline
\end{tabular}
\caption{PGF test results for 3+1 and our model, where Eqs.~\ref{chi2pg} and \ref{npg} explain how $\chi^2_{PGF}$ and $N_{PGF}$ are determined.}
\label{table:fitquality}
\end{table*}

\end{document}